\documentclass[12pt,reqno,a4paper]{article}
\usepackage[utf8]{inputenc}
\usepackage[british]{babel}
\usepackage{graphicx} %Per a poder incloure imatges
\usepackage{amsmath}   %  Paquets per
\usepackage{amssymb}   %  matemàtics
\usepackage{amsfonts}
\usepackage{systeme,mathtools}
\usepackage{booktabs}
\usepackage{enumerate}
\usepackage{subcaption}
\usepackage{url}
\usepackage{mathdots}
\usepackage{multicol}
\usepackage{fancyhdr}
\usepackage{xcolor}
\usepackage{tikz-cd}

\newtheorem{theorem}{Theorem}

\newtheorem{definition}{Definition}

\newcommand{\bbC}{{\mathbb C}}
\newcommand{\bbP}{{\mathbb P}}

\newcommand{\ka}{{\kappa}}

\def\t{\widetilde}

\def\ja#1{{\color{orange}{#1}}} % Jaume's colour

\title{Discrete Painlev\'e equations  from pencils of quadrics\\ in $\bbP^3$ with branching generators}
\author{Jaume Alonso, Yuri B. Suris}
\date{\small Institut für Mathematik, MA 7-1\\ Technische Universität Berlin, Str.\ des 17.\ Juni, 10623 Berlin, Germany\\
E-mail: alonso@math.tu-berlin.de, suris@math.tu-berlin.de}

\begin{document}

\maketitle

{\bf Abstract.} In this paper we extend the novel approach to discrete Painlev\'e equations initiated in our previous work \cite{ASWpart1}.  A classification scheme for discrete Painlev\'e equations proposed by Sakai interprets them as birational isomorphisms between generalized Halphen surfaces (surfaces obtained from $\mathbb P^1\times\mathbb P^1$ by blowing up at eight points). Sakai's classification is thus based on the classification of generalized Halphen surfaces. In our scheme, the family of generalized Halphen surfaces is replaced by a pencil of quadrics in $\mathbb P^3$. A discrete Painlev\'e equation is viewed as an autonomous transformation of $\mathbb P^3$ that preserves the pencil and maps each quadric of the pencil to a different one. Thus, our scheme is based on the classification of pencils of quadrics in $\mathbb P^3$. Compared to our previous work, here we consider a technically more demanding case where the characteristic polynomial $\Delta(\lambda)$ of the pencil of quadrics is not a complete square. As a consequence, traversing the pencil via a 3D Painlev\'e map corresponds to a translation on the universal cover of the Riemann surface of $\sqrt{\Delta(\lambda)}$, rather than to a M\"obius transformation of the pencil parameter $\lambda$ as in \cite{ASWpart1}.

\section{Introduction}
\label{sect intro}

This paper is the second contribution to our study devoted to a novel interpretation of discrete Painlev\'e equations, which builds up on \cite{ASWpart1}. Discrete Painlev\'e equations belong to the most intriguing objects in the theory of discrete integrable systems. After some examples sporadically appeared in various applications, their systematic study started when Grammaticos, Ramani and Papageorgiou proposed the notion of ``singularity confinement'' as an integrability detector, and found the first examples of second order nonlinear non-autonomous difference equations with this property, which they denoted as \emph{discrete Painlev\'e equations} \cite{GRP, RGH}. The activity of their group was summarized in \cite{GRreview}. A general classification scheme of discrete Painlev\'e equations was proposed by Sakai \cite{Sakai} and it is given a detailed exposition in the review paper by Kajiwara, Noumi and Yamada \cite{KNY}. In the framework of Sakai's scheme, discrete Painlev\'e equations are birational maps between \emph{generalized Halphen surfaces} $X$. The latter can be realized as $\bbP^1\times\bbP^1$ blown up at eight points.
A monographic exposition of discrete Painlev\'e equations is given by Joshi \cite{J}.
\smallskip

Let us summarize the main ingredients and features of our alternative approach to discrete Painlev\'e equations, initiated in \cite{ASWpart1}. 
\begin{itemize}
\item A pencil of quadrics $\{Q_\lambda\}$ in $\mathbb P^3$ containing non-degenarate quadrics. Such pencils can be classified modulo projective transformations of $\mathbb P^3$, and they come in thirteen classes. The class of the pencil can be identified by the type of its {\em base curve} $Q_0\cap Q_\infty$. This is a spatial curve of degree 4, whose type can vary from a generic one (irreducible smooth curve for a pencil of type (i)), through irreducible curves with a node (type (ii)) or with a cusp (type (iii)), to various types of reducible curves (from two non-coplanar conics intersecting at two points, type (iv), to a pair of intersecting double lines, type (xiii)).

\item The second pencil of quadrics $\{P_\mu\}$ having one quadric in common with $\{Q_\lambda\}$, say $P_\infty=Q_\infty$. The base curves of both pencils intersect at eight points $S_i$, $i=1,\ldots,8$. 

\item Given two pencils of quadrics, one can define a three-dimensional analog of a QRT map $F=i_1\circ i_2$, where the 3D QRT involutions $i_1$, $i_2$ act along two families of generators of $Q_\lambda$, see \cite{ASW}. Each involution puts into correspondence two intersection points of a generator with the quadric $P_\mu$. By definition, such an involution, and therefore the 3D QRT map $F=i_1\circ i_2$, leaves each quadric of two pencils invariant, and thus possesses two rational integrals of motion $\lambda=Q_0/Q_\infty$ and $\mu=P_0/P_\infty$.

\item A {\em Painlev\'e deformation map} is the device which allows us to travel across the pencil $\{Q_\lambda\}$. More precisely, such a map $L$ on $\mathbb P^3$ preserves the pencil, but not fiber-wise. Rather, it sends each quadric $Q_\lambda$ to a different quadric $Q_{\widehat\lambda}$. Moreover, $L$ preserves the base curve of the pencil $\{Q_\lambda\}$. In the cases considered in \cite{ASWpart1}, the base curve is reducible and contains straight lines. In these cases, $L$ does not necessarily fix these straight lines point-wise. In the cases considered in the present paper, $L$ fixes the base curve $Q_0\cap Q_\infty$ pointwise (in particular, it fixes all eight points $S_i$).

\item A {\em 3D Painlev\'e map} is obtained by composition $\t F=L\circ i_1\circ L\circ i_2$, provided it possesses the singularity confinement property. It is to be stressed that the pencil $\{Q_\lambda\}$ continues to play a fundamental role in the dynamics of $\t F$: the maps $L\circ i_1$, $L\circ i_2$ preserve the pencil and map each quadric $Q_\lambda$ to $Q_{\widehat \lambda}$. We do not have a straightforward description of the dynamical role of the pencil $\{P_\mu\}$, but anticipate its relation to the isomonodromic description of the discrete Painlev\'e equations.
\end{itemize}

One can say that in our approach the role of a family of generalized Halphen surfaces is played by the quadrics of the pencil $\{Q_\lambda\}$ with eight distinguished points on the base curve of the pencil. The base curve itself plays the role of the unique anti-canonical divisor. Let us stress several features of our construction which are in a sharp contrast to the Sakai scheme.
\begin{itemize}
\item Neither the exceptional divisor nor the eight distinguished points evolve under the map $\t F$. Their discrete time evolution is apparent and is due to their representation in the so-called {\em pencil-adapted coordinates}. These are coordinates $(x,y,\lambda)\in\mathbb P^1\times\mathbb P^1\times\mathbb P^1$ establishing an isomorphism between each quadric $Q_\lambda$ of the pencil and $\mathbb P^1\times \mathbb P^1$. The pencil-adapted coordinates of a point on the base curve do depend on $\lambda$, so traversing the pencil $\lambda\mapsto \widehat\lambda$ under $\t F$ induces an {\em apparent discrete time evolution} of the base curve and of the eight distinguished points.

\item The shift parameter $\delta$ of discrete Painlev\'e equations (or its exponent $q=e^\delta$ for the $q$-difference equations among them) is not an intrinsic characteristic of the configuration of eight distinguished points, but is a free parameter of the construction.
\end{itemize}

One can say that our approach is a realization of the old-style idea of discrete Painlev\'e equations being non-autonomous versions (or modifications) of the QRT maps. This idea was instrumental in the discovery and early classification attempts of discrete Painlev\'e equations, summarized in \cite{GRreview}. A more geometric version of this procedure was proposed in the framework of the Sakai's scheme by Carstea, Dzhamay and Takenawa \cite{CDT}. In their scheme, the de-autonomization of a given QRT map depends on the choice of one biquadratic curve of the pencil. In our approach, the choice of the base curve and eight distinguished point on it determines uniquely all the ingredients of the construction, starting with the \ja{two} pencils of quadrics.

The structure of the paper is as follows. In Section \ref{sect gen}, we describe the construction scheme of discrete Painlev\'e equations applicable to the present case and stress its distinctions from the previous paper \cite{ASWpart1}. The main distinction is that here we consider the pencils whose characteristic polynomial $\Delta(\lambda)$ is not a complete square. As a consequence, the 3D QRT involutions $i_1$, $i_2$ and the 3D QRT map $F=i_1\circ i_2$ are no more birational maps of $\mathbb P^3$. Rather, these maps become birational maps on $\mathcal X$, a branched double covering of $\bbP^3$, whose ramification locus is the union of the singular quadrics $Q_{\lambda_i}$, where $\lambda_i$ are the branch points of the Riemann surface $\mathcal R$ of $\sqrt{\Delta(\lambda)}$. 

In Section \ref{sect L}, we formulate a general recipe for the construction of the Painlev\'e deformation map $L$, responsible to the evolution $\lambda\mapsto\widehat\lambda$ across the pencil of quadrics $\{Q_\lambda\}$. While in the first part \cite{ASWpart1} we had $\widehat\lambda=\sigma(\lambda)$,  where $\sigma:\bbP^1\to\bbP^1$ is a M\"obius automorphism fixing the set 
${\rm Sing}(Q):=\big\{\lambda\in\bbP^1: \;Q_\lambda\;{\rm is\;\; degenerate}\big\}$, in the present paper the natural definition becomes $\widehat\lambda=\lambda(\widehat\nu)$, where $\lambda=\lambda(\nu)$ is the holomorphic uniformization map for the Riemann surface $\mathcal R$, and $\widehat\nu=\nu+2\delta$ is the translation on the universal cover $\mathbb C$. The recipe turns out to be applicable to all types of the pencil $\{Q_\lambda\}$ except for the generic type (i). The latter leads to the elliptic Painlev\'e equation, which will be treated in a separate publication.

In Section \ref{sect conf}, we show that the so constructed $L$ ensures the fundamental singularity confinement property for our 3D Painlev\'e maps.

There follow five Sections \ref{sect dPE7}--\ref{sect qPE8} containing a detailed elaboration of our scheme for all relevant types of the pencils except for the type (i). We recover, within our novel framework, all discrete Painlev\'e equations except for the elliptic one, which is left for a separate publication.

\subparagraph*{Acknowledgement.}
This research was supported (till June 2024) by the DFG Collaborative Research Center TRR 109 ``Discretization in
Geometry and Dynamics''.

%%%%%%%%%%%%%%%%%%%%%%%%%%%%%%%
\section{General scheme}
\label{sect gen}
%%%%%%%%%%%%%%%%%%%%%%%%%%%%%%%

We now describe the construction scheme of discrete Painlev\'e equations applicable to the present case and stress its distinctions from the previous paper \cite{ASWpart1}. The first steps are the same as there:
\begin{itemize}
\item Start with a pencil $\{C_\mu\}$ of biquadratic curves in $\bbP^1\times\bbP^1$ and the corresponding QRT map. Let $s_1,\ldots,s_8\in\bbP^1\times\bbP^1$ be the base points of this pencil. Lift $\{C_\mu\}$ to a pencil of quadrics $\{P_\mu\}$ in $\bbP^3$ using the \emph{Segre embedding} of $\bbP^1\times\bbP^1$ to $\bbP^3$. The base curve of this pencil passes through the lifts $S_1,\ldots,S_8$ of the base points $s_1,\ldots,s_8$. 
\item Choose one distinguished biquadratic curve $C_\infty$ of the pencil, along with its lift to a quadric $P_\infty$.
\item Based on these data, construct the pencil of quadrics $\{Q_\lambda=Q_0-\lambda Q_\infty\}$  in $\bbP^3$ spanned by $Q_0=\{X_1X_2-X_3X_4=0\}$ and $Q_\infty:=P_\infty$. Recall that $Q_0$ is nothing but the 
image of $\bbP^1\times\bbP^1$ by the Segre embedding. The base curve of the pencil $\{Q_\lambda\}$ is, by definition, the curve $Q_0\cap Q_\infty$, which is the image of $C_\infty$ under the Segre embedding. The intersection of this curve with the base curve of the pencil $\{P_\mu\}$ consists exactly of the points $S_1,\ldots,S_8$.
\end{itemize}
The characteristic polynomial of the pencil $\{Q_\lambda\}$ is
\begin{equation}
\Delta(\lambda)=\det(M_\lambda)=\det(M_0-\lambda M_\infty),
\end{equation}
where $M_0,M_\infty\in {\rm Sym}_{4\times 4}(\bbC)$ are symmetric matrices of the quadratic forms $Q_0,Q_\infty$. In the present paper, we are dealing with the cases where this polynomial is {\em not a complete square}. According to the projective classification of pencils of quadrics, discussed in \cite{ASWpart1}, these are the following six cases:
\begin{itemize}

\item[(i)] \emph{Pencil of quadrics through a non-singular spatial quartic curve.}
\newline
Segre symbol $[1,1,1,1]$; $\Delta(\lambda)=(\lambda-\lambda_1)(\lambda-\lambda_2)(\lambda-\lambda_3)(\lambda-\lambda_4)$.

\item[(ii)] \emph{Pencil of quadrics through a nodal spatial quartic curve}.
\newline
Segre symbol $[2,1,1]$;  $\Delta(\lambda)=(\lambda-\lambda_1)^2(\lambda-\lambda_2)(\lambda-\lambda_3)$.

\item[(iii)] \emph{Pencil of quadrics through a cuspidal spatial quartic curve}. 
\newline 
Segre symbol $[3,1]$; $\Delta(\lambda)=(\lambda-\lambda_1)^3(\lambda-\lambda_2)$.

\item[(iv)] \emph{Pencil of quadrics through two non-coplanar conics sharing two points.} 
\newline 
Segre symbol $[(1,1),1,1]$; $\Delta(\lambda)=(\lambda-\lambda_1)^2(\lambda-\lambda_2)(\lambda-\lambda_3)$.

\item[(v)] \emph{Pencil of quadrics through two non-coplanar conics touching at a point.}
\newline 
Segre symbol $[(2,1),1]$; $\Delta(\lambda)=(\lambda-\lambda_1)^3(\lambda-\lambda_2)$.

\item[(vi)] \emph{Pencil of quadrics tangent along a non-degenerate conic.}
\newline 
Segre symbol  $[(1,1,1),1]$; $\Delta(\lambda)=(\lambda-\lambda_1)^3(\lambda-\lambda_2)$.
\end{itemize}

As discussed in \cite{ASWpart1}, for $X\in Q_\lambda$, the generators $\ell_1(X)$ and $\ell_2(X)$ are rational functions of $X$ and of $\sqrt{\Delta(\lambda)}$. The dependence on $\lambda$ can be expressed as a holomorphic dependence on the point of the Riemann surface $\mathcal R$ of $\sqrt{\Delta(\lambda)}$. This Riemann surface is a double cover of $\widehat{\mathbb C}$ branched at two or at four points. By the uniformization theorem, its universal cover is $\mathbb C$. We will denote the uniformizing variable $\nu\in\mathbb C$, so that the maps $\nu\mapsto \lambda$ and $\nu\mapsto \sqrt{\Delta(\lambda)}$ are holomorphic. The following three situations can be distinguished:
\begin{itemize}
\item[-] case (i): four distinct branch points $\lambda_1,\lambda_2,\lambda_3,\lambda_4$, the Riemann surface $\mathcal R$ is a torus, whose conformal class is determined by the cross-ratio of the branch points. This case, corresponding to the elliptic Painlev\'e equations, will be treated in an upcoming work;
\item[-] cases (ii), (iv): two branch points $\lambda_2,\lambda_3$, one of the periods of the torus becomes infinite, so that $\mathcal R$ is a cylinder;
\item[-] cases (iii), (v), (vi): two branch points $\lambda_1,\lambda_2$, both periods of the torus become infinite, so that $\mathcal R$ is plane.
\end{itemize}
It becomes necessary to introduce modifications in the two major ingredients of the construction in \cite{ASWpart1}.
\begin{itemize}
\item The generators $\ell_1$, $\ell_2$ are not rational functions on $\mathbb P^3$ anymore. Rather, they become well-defined rational maps on the variety $\mathcal X$ which is a branched double covering of $\bbP^3$, whose ramification locus is the union of the singular quadrics $Q_{\lambda_i}$, where $\lambda_i$ are the branch points of $\mathcal R$. The same is true for a linear projective change of variables $X=A_\nu Y$ reducing the quadratic form $Q_{\lambda(\nu)}$ to the standard form $Q_0$, which we now write as
\begin{equation}\label{A norm}
Q_{\lambda(\nu)}(A_{\nu} Y)=Q_0(Y), \quad {\rm or}\quad A_{\nu}^{\rm T}M_{\lambda(\nu)} A_{\nu}=M_0,
\end{equation}
and for the \emph{pencil-adapted coordinates} 
\begin{equation}\label{phi lambda}
\begin{bmatrix} X_1 \\ X_2 \\ X_3 \\ X_4 \end{bmatrix}=A_{\nu}\begin{bmatrix} x \\ y \\ xy \\ 1 \end{bmatrix} =:\phi_{\nu}(x,y).
\end{equation} 
Thus, $\phi_{\nu}$ gives a parametrization of $Q_{\lambda(\nu)}$ by $(x,y)\in\bbP^1\times\bbP^1$, such that the generators $\ell_1$, resp.\ $\ell_2$ of $Q_\lambda$ correspond to $x={\rm const}$, resp.\ to $y={\rm const}$. Interchanging two sheets of the covering corresponds to interchanging two families of generators $\ell_1$, $\ell_2$.

\item Also the \emph{3D QRT involutions} $i_1$, $i_2$ for the pencil $\{Q_\lambda\}$, defined by intersections of its generators $\ell_1$, $\ell_2$ with the quadrics $P_\mu$ (see \cite{ASW}), are not birational maps of $\mathbb P^3$ anymore, and the same is true for the {\em 3D QRT map} $F=i_1\circ i_2$. Rather, these maps become birational maps on $\mathcal X$.
\end{itemize}
The next main deviation from the construction of \cite{ASWpart1} is that it becomes unnatural to consider Painlev\'e deformation maps $L$ as birational maps $\bbP^3$ preserving the pencil $\{Q_\lambda\}$ and sending each $Q_\lambda$ to $Q_{\sigma(\lambda)}$, where $\sigma:\bbP^1\to\bbP^1$ is a M\"obius automorphism fixing the set 
$
{\rm Sing}(Q):=\{\lambda\in\bbP^1: Q_\lambda\;{\rm is\; degenerate}\}.
$
Instead, in the present context we formulate the following requirement.
\begin{itemize}
\item A Painlev\'e deformation map $L$ is a birational map on $\mathcal X$ preserving the pencil $\{Q_\lambda\}$ and sending $Q_{\lambda(\nu)}$ to $Q_{\lambda(\widehat\nu)}$, where $\nu\mapsto\widehat\nu=\nu+2\delta$ is a translation on the universal cover of $\mathcal R$.
\end{itemize}

As compared with \cite{ASWpart1}, our construction will involve some additional ingredients, required to establish the relation to the form of discrete Painlev\'e equations known from the literature. The Painlev\'e deformation map $L$ is decomposed in two factors, each one depending only on one of the variables $x,y$, and shifting the variable $\nu$ by $\delta$. This can be done in two ways:
\begin{equation}\label{L fact 1}
L=L_1\circ R_2, \;\; {\rm where}\; \;L_1:(x,y,\nu)\mapsto (x,\widetilde y,\nu+\delta), \;\; R_2:(x,y,\nu)\mapsto (\widetilde x,y,\nu+\delta),
\end{equation} 
resp.\ 
\begin{equation}\label{L fact 2}
L=L_2\circ R_1, \;\;{\rm where}\;\; L_2:(x,y,\nu)\mapsto (\widetilde x,y,\nu+\delta), \;\; R_1:(x,y,\nu)\mapsto (x,\widetilde y,\nu+\delta).
\end{equation} 
(The indices $1$, $2$ refer to the variables which {\em do not} change under the map, like for $i_1$, $i_2$.)
Each one of $L_1,L_2,R_1,R_2$ maps $Q_{\lambda(\nu)}$ to $Q_{\lambda(\nu+\delta)}$. We set 
$$
\nu_n=\nu_0+2n\delta \quad {\rm for}\quad n\in \frac{1}{2}\mathbb Z,
$$ 
so that $\nu_{n+1/2}=\nu+\delta$. The variables associated to the discrete Painlev\'e equations known from the literature, parametrize in our formulation the quadrics with half-integer indices, namely
$$
(x_n,y_n,\nu_{2n-1/2})\in Q_{\lambda(\nu_{2n-1/2})}\ , \quad (x_{n+1},y_n,\nu_{2n+1/2})\in Q_{\lambda(\nu_{2n+1/2})}\ .
$$
\begin{definition}
A 3D Painlev\'e map is given by 
\begin{equation}
\t F=\t i_1\circ \t i_2, \quad{where}\quad \t i_1=R_1\circ i_1\circ L_1, \quad \t i_2= R_2\circ i_2\circ L_2,
\end{equation} 
or, in coordinates,
\begin{align}
& (x_n,y_n,\nu_{2n-1/2})\stackrel{L_2}{\to} (x,y_n,\nu_{2n}) \stackrel{i_2}{\to} (\t x, y_n,\nu_{2n})\stackrel{R_2}{\to} (x_{n+1},y_n,\nu_{2n+1/2}) \label{dP def eq 1}\\
& \qquad \stackrel{L_1}{\to}(x_{n+1}, y,\nu_{2n+1})  \stackrel{i_1}{\to} (x_{n+1},\t y,\nu_{2n+1}) \stackrel{R_1}{\to}(x_{n+1},y_{n+1},\nu_{2n+3/2}). \label{dP def eq 2}
\end{align}
\end{definition}
The map $\t F$ is conjugate to $L\circ i_1\circ L\circ i_2$; note that the latter map acts between the quadrics with integer indices.
\smallskip

Our last requirement repeats the one in \cite{ASWpart1}:
\begin{itemize}
\item The singularity confinement properties of $\tilde i_1$, $\tilde i_2$ are the same as that of $i_1$, $i_2$.
\end{itemize}

\paragraph{Reduction to the symmetric case.} If the eight points $s_i$ are symmetric with respect to the symmetry switch $\sigma:(x,y)\mapsto (y,x)$, we can define a 2D QRT root $f=i_1\circ\sigma=\sigma\circ i_2$ such that $F=f\circ f$. In this case, the map $L$ in the pencil-adapted coordinates satisfies $L=\sigma\circ L\circ \sigma$, and therefore its decomposition factors satisfy 
$$
L_2=\sigma \circ L_1\circ \sigma, \quad R_2=\sigma \circ R_1\circ \sigma.
$$
The 3D Painlev\'e map $F$ can be written as 
\begin{eqnarray*}    
    \t F &=& R_1\circ i_1\circ L_1\circ R_2 \circ i_2\circ L_2=
     R_1\circ i_1\circ L_1\circ \sigma \circ R_1 \circ \sigma \circ i_2\circ \sigma \circ L_1 \\
        &=&   R_1\circ i_1\circ \sigma \circ L_2\circ R_1 \circ \sigma \circ i_2\circ L_2 \\
        &=&   (R_1 \circ f\circ  L_2)^2.
\end{eqnarray*}
Therefore, one can define the Painlev\'e deformed QRT root as $\t f=R_1\circ f\circ L_2$, then the discrete Painlev\'e map decomposes as $\t F=\t f\circ\t f$.

%%%%%%%%%%%%%%%%%%%%%%%%%%%%%%
\section{Construction of the Painlev\'e deformation map}
\label{sect L}
%%%%%%%%%%%%%%%%%%%%%%%%%%%%%%

The desired properties of the Painlev\'e deformation map $L$ are ensured by the following construction.

\begin{theorem}\label{th L}
If the polynomial $Q_\infty$ does not depend on $X_3$, define the map $L:[X_1:X_2:X_3:X_4]\mapsto [\widehat X_1:\widehat X_2:\widehat X_3:\widehat X_4]$ by requiring that, for $X\in Q_{\lambda(\nu)}$, there holds
\begin{equation}\label{eq L 1st version}
\left\{
\begin{array}{l}
\widehat X_1=X_1X_4,\\
\widehat X_2=X_2X_4,\\
 \widehat X_3=X_3X_4-\big(\lambda(\widehat{\nu})-\lambda(\nu)\big)Q_\infty(X),\\
 \widehat X_4=X_4^2,
\end{array}
\right.
\end{equation}
where $\widehat\nu=\nu+2\delta$. If $Q_\infty$ does not depend on $X_1$, define
\begin{equation}\label{eq L 2nd version}
\left\{
\begin{array}{l}
\widehat X_1 = X_1X_2+\big(\lambda(\widehat{\nu})-\lambda(\nu)\big)Q_\infty(X),\\
\widehat  X_2=X_2^2,\\
\widehat  X_3=X_2X_3,\\
\widehat  X_4=X_2X_4.
\end{array}
\right.
\end{equation}
Then $L$ sends each $Q_{\lambda(\nu)}$ to $Q_{\lambda(\widehat\nu)}$ and fixes all points of the base curve of the pencil $\{Q_\lambda\}$ not belonging to $\{X_4=0\}$ (resp.\ to $\{X_2=0\}$), including all eight base points $S_i$, $i=1,\ldots,8$.
\end{theorem}
{\bf Proof.} It follows by a simple computation. For instance, for the case \eqref{eq L 1st version}:
$$
\widehat X_1\widehat X_2-\widehat X_3\widehat X_4-\lambda(\widehat\nu)Q_\infty(\widehat X)=
X_4^2\Big(X_1X_2-X_3X_4-\lambda(\nu)Q_\infty(X)\Big).
$$
Futher, if $Q_\infty(X)=0$ and $X_4\neq 0$, then $[\widehat X_1:\widehat X_2:\widehat X_3:\widehat X_4]=[X_1:X_2:X_3:X_4]$. $\blacksquare$
\smallskip

The recipe of Theorem \ref{th L} covers all cases treated in the present paper (pencils of the types (ii)-(vi)). In retrospect, we notice that, with a natural modification (replace $\widehat\lambda-\lambda=\lambda(\widehat\nu)-\lambda(\nu)$ by $\sigma(\lambda)-\lambda$), this recipe covers also the cases considered in the first part of this study \cite{ASWpart1}. For pencils of the type (i) the quadric $Q_\infty$ is non-degenerate, so a modification of the recipe is required.

%%%%%%%%%%%%%%%%%%%%%%%%
\section{Singularity confinement}
\label{sect conf}
%%%%%%%%%%%%%%%%%%%%%%%%

Our case-by-case computations reveal the following observation. In all examples of the present paper,  the eight points $s_1,\ldots,s_8$ in $\bbP^1\times \bbP^1$ serve as the  indeterminacy set for the 2D QRT involutions  $i_1$, $i_2$. The singularity confinement structure can be summarised as:
\begin{equation}\label{sing1}
 \{x=a_i\}\;\xrightarrow{i_1}\; s_i\;\xrightarrow{i_2}\; \{y=b_i\}, \quad i=1,\ldots,8.
\end{equation}
In the pencil-adapted coordinates, the 3D QRT involutions restricted to $Q_{\lambda(\nu)}$ are given by the same formulas as the original 2D QRT involutions, with the points $s_i$ replaced by their deformations $s_i(\nu)$. The latter still support a pencil of biquadratic curves, which are the pre-images under $\phi_\nu$ of the intersection curves $Q_{\lambda(\nu)}\cap P_\mu$. Therefore, for the 3D QRT involutions $i_1$ and $i_2$, we have 
\begin{equation}\label{s nu}
\{x=a_i(\nu)\} \;\stackrel{i_1}{\to}\; s_i(\nu)\;\stackrel{i_2}{\to}\;\{y=b_i(\nu)\} .
 \end{equation}
 Let $\Phi_i\subset\bbP^3$ be the ruled surface comprised of the lines on $Q_{\lambda(\nu)}$ given, in the pencil-adapted coordinates $\phi_\nu$, by the equations $\{x=a_i(\nu)\}$. Likewise, let $\Psi_i\subset\bbP^3$ be the ruled surface comprised of the lines on $Q_{\lambda(\nu)}$ given in the coordinates $\phi_\nu$ by the equations $\{y=b_i(\nu)\}$. Then, in view of \eqref{s nu}, we obtain the following singularity confinement patterns for $i_1, i_2$:
\begin{equation}\label{sing conf 3D QRT i1i2}
\Phi_i\; \stackrel{i_1}{\rightarrow}\; S_i\; \stackrel{i_2}{\rightarrow}\; \Psi_i.
\end{equation} 
We emphasize that the surfaces $\Phi_i$ are blown down to points (rather than curves), and these points are blown up to surfaces $\Psi_i$ again.
\begin{theorem}\label{prop Painleve}
Suppose that the involutions $i_1, i_2:\bbP^3\dasharrow\bbP^3$ have a singularity confinement pattern of the type \eqref{sing conf 3D QRT i1i2}, and $L$ satisfies
\begin{equation}\label{cond 1}
L(S_i)=S_i.
\end{equation} 
Then for the deformed maps $\t i_1=R_1\circ i_1\circ L_1$, $\t i_2=R_2\circ i_2\circ L_2$ we have:
\begin{equation}\label{sing conf dP i1i2}
L_1^{-1}(\Phi_i)\; \stackrel{\t i_1}{\rightarrow}\; R_1(S_i)\; \stackrel{\t i_2}{\rightarrow}\; R_2(\Psi_i),
\end{equation}
which implies for $\t F=\t i_1\circ \t i_2$ the singularity confinement pattern
\begin{equation}\label{sing conf dP f short}
(L_1\circ \t i_2)^{-1}(\Phi_i)\; \stackrel{\t F}{\rightarrow}\; R_1(S_i)\; \stackrel{\t F}{\rightarrow}\; (\t i_1\circ R_2)(\Psi_i).
\end{equation}

\end{theorem}

An important observation is that the eight points $R_1(S_i)$ participating in these singularity confinement patterns do not support a net of quadrics.

%%%%%%%%%%%%%%%%%%%%%%%%%%%%%%%
%%%%%%%%%%%%%%%%%%%%%%%%%%%%%%%
\section{From a pencil of type (v) to the \textit{d}-Painlev\'e equation of the surface type $A_1^{(1)}$}
\label{sect dPE7}
%%%%%%%%%%%%%%%%%%%%%%%%%%%%%%%
%%%%%%%%%%%%%%%%%%%%%%%%%%%%%%%

\paragraph{2D QRT map.}
We consider the QRT map corresponding to the pencil of biquadratic curves $\{C_\mu\}$ through eight points $s_i=(a_i,b_i)$, $i=1,\ldots,8$, where
\begin{equation}
b_i=-a_i, \;\; i=1,\ldots,4, \quad {\rm and}\quad b_i=1-a_i,\;\; i=5,\ldots,8. 
\end{equation}
These eight points support a pencil of biquadratic curves if they satisfy the condition 
\begin{equation}\label{dPE7 cond}
   \sum_{i=1}^8 a_i=\sum_{i=1}^8b_i\;\;\Leftrightarrow\;\; \sum_{i=1}^8 a_i=2.
\end{equation}
This pencil contains a reducible curve, consisting of two (1,1)-curves: 
\begin{equation}
C_\infty=\{(x+y)(x+y-1)=0\}.
\end{equation}

%%%%%%%%%%%%%%%%%%%%%%%%%%%%%%%%%
\begin{figure}[!ht]
\begin{center}
\begin{subfigure}[b]{0.3\textwidth}
         \centering
         \includegraphics[width=0.8\textwidth]{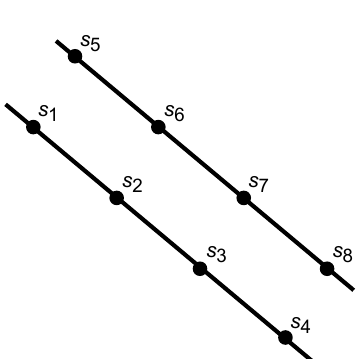}
         \caption{}
     \end{subfigure}
     \hspace{2cm}
     \begin{subfigure}[b]{0.3\textwidth}
         \centering
         \includegraphics[width=\textwidth]{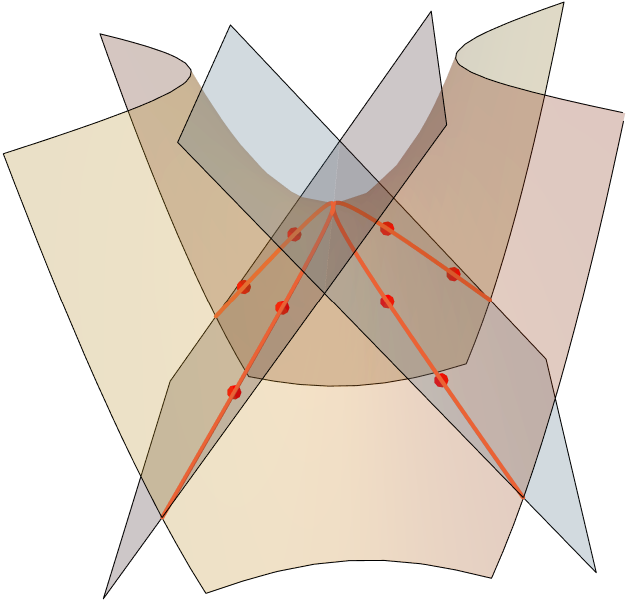}
         \caption{}
     \end{subfigure}
\end{center}
\caption{(a) Base set of the surface type $A_1^{(1)}$: two quadruples of points on two touching (1,1)-curves in $\bbP^1\times\bbP^1$. (b) Pencil of quadrics through two touching non-coplanar conics}
\label{fig dPE7}
\end{figure}
%%%%%%%%%%%%%%%%%%%%%%%%%%%%%%%%%

The vertical involution $i_1$ for this pencil can be described by the following equation:
 \begin{equation} \label{dPE7 QRT i1}
 i_1(x,y)=(x,\t y),\quad  \frac{(\t y+x)(x+y)}{(\t y+x-1)(x+y-1)}=\frac{\prod_{i=1}^{4}(x-a_i)}{\prod_{i=5}^{8}(x-a_i)}.
 \end{equation}
Similarly, the horizontal involution $i_2$ can be described by the following equation:
 \begin{equation} \label{dPE7 QRT i2}
i_2(x,y)=(\t x,y),\quad   \frac{(\t x+y)(x+y)}{(\t x+y-1)(x+y-1)}=\frac{\prod_{i=1}^{4}(y+a_i)}{\prod_{i=5}^{8}(y+a_i-1)}.
 \end{equation}
The QRT map is the composition of these two involutions, $F=i_1\circ i_2$. The singularity confinement structure of the involutions $i_1$, $i_2$ is as in \eqref{sing1}. The symmetric case corresponds to
$$
a_{2i}=-a_{2i-1}, \; i=1,2, \qquad a_{2i}=1-a_{2i-1}, \, i=3,4.
$$
In this case, $F=f\circ f$, with  $f=i_1\circ \sigma=\sigma\circ i_2$ being the QRT root (here $\sigma(x,y)=(y,x)$).

\paragraph{3D Painlev\'e map.} We consider the pencil of quadrics $\{P_\mu\}$ in $\bbP^3$, the Segre lift of the pencil of curves $\{C_\mu\}$. The pencil $\{Q_\lambda\}$ is spanned by $Q_0=\{X_1X_2-X_3X_4=0\}$ and
$Q_\infty=P_{\infty}=\{(X_1+X_2) (X_1+X_2-X_4)=0\}$:
\begin{equation}\label{dPE7 pencil}
Q_\lambda=\Big\{X_1X_2-X_3X_4-\lambda(X_1+X_2) (X_1+X_2-X_4)=0\Big\}.
\end{equation}
The base set of the pencil $Q_\lambda$ consists of the two conics, $\{X_1X_2-X_3X_4=0,\; X_1+X_2=0\}$ and  $\{X_1X_2-X_3X_4=0,\; X_1+X_2=X_4\}$, which have one common (touching) point $[0:0:1:0]$. This is a pencil of type (v). The intersection of this base set with the base set of the pencil $\{P_\mu\}$ consists of eight points
\begin{equation*}
S_i=[a_i:b_i:a_ib_i:1], \quad i=1,\ldots,8,
\end{equation*}
which are nothing but  the lifts of the points $s_i$ under the Segre embedding. 

The matrix $M_\lambda$ of the quadratic form $Q_\lambda$:
\begin{equation}
M_\lambda=\begin{pmatrix}
-2\lambda & 1-2\lambda & 0  & -\lambda\\
1-2\lambda & -2\lambda & 0 & -\lambda \\
0 & 0 & 0 & -1 \\
-\lambda & -\lambda & -1 & 0 
\end{pmatrix}.
\end{equation}
The characteristic polynomial of the pencil $\{Q_\lambda\}$ is: $\Delta(\lambda)=\det(M_\lambda)=-1+4\lambda$, which is not a complete square, and ${\rm Sing}(Q_\lambda)=\{\frac{1}{4},\infty\}$. We uniformize the Riemann surface of $\sqrt{\Delta(\lambda)}$ via
\begin{equation}\label{dPE7 unif}
\lambda=\frac{1-\nu^2}{4}, \quad  \sqrt{\Delta(\lambda)}=\nu.
\end{equation}
Thus, $\lambda(-\nu)=\lambda(\nu)$, while $\sqrt{\Delta(\lambda)}$ changes its sign as $\nu\to-\nu$. This gives us a double cover of the original pencil branched at $\lambda=1/4$ (corresponding to $\nu=0$), and at $\lambda=\infty$ (corresponding to $\nu=\infty$). The normalizing transformation of $Q_{\lambda(\nu)}$ to the canonical form $Q_0$ can be found as follows: 
\begin{equation}\label{dPE7 pencil norm}
\begin{bmatrix} X_1\\X_2\\X_3\\X_4\end{bmatrix} = A_\nu \! \begin{bmatrix} Y_1 \\ Y_2 \\ Y_3 \\ Y_4\end{bmatrix}, 
\end{equation}
where
\begin{equation}
A_\nu=\begin{pmatrix} \frac{1}{2\nu}(1+\nu) & \frac{1}{2\nu}(1-\nu) & 0 &  0 \\[0.15cm] 
\frac{1}{2\nu}(1-\nu) & \frac{1}{2\nu}(1+\nu) & 0 & 0 \\[0.15cm]  
\frac{1}{4\nu}(1-\nu^2) & \frac{1}{4\nu}(1-\nu^2) & 1 & 0 \\[0.15cm] 
0 & 0 & 0 & 1 \end{pmatrix}.                               
 \end{equation}
 Indeed, one immediately verifies that 
 $$
 A_\nu^{\rm T} M_{\lambda(\nu)} A_\nu=M_0.
 $$
Now, we are in the position to derive a parametrization of the quadric $Q_\lambda$:
\begin{equation}\label{dPE7 x to X}
\begin{bmatrix} X_1 \\ X_2 \\ X_3 \\ X_4 \end{bmatrix}=A_\nu \begin{bmatrix} x \\ y \\ xy \\ 1 \end{bmatrix}
=\begin{bmatrix} \frac{1}{2\nu}((1+\nu)x+(1-\nu)y) \\[0.15cm] 
\frac{1}{2\nu}((1-\nu)x+(1+\nu)y) \\[0.15cm] 
xy+\frac{1-\nu^2}{4\nu}(x+y) \\[0.15cm] 1  \end{bmatrix}=:\phi_\nu(x,y).
\end{equation}
Observe that this parametrization is neither valid for $\nu=0$ nor for $\nu=\infty$.
The pencil-adapted coordinates $(x,y,\nu)$ on (the double cover of) $\bbP^3$ are:
\begin{equation}\label{dPE7 X to x}
x = \frac{(1+\nu)X_1-(1-\nu)X_2}{2X_4}, \quad y = \frac{(1+\nu)X_2-(1-\nu)X_1}{2X_4},
\end{equation}
which have to be supplemented with
\begin{equation}\label{dPE7 X to lambda}
\lambda=\frac{1-\nu^2}{4}=\frac{X_1X_2-X_3X_4}{(X_1+X_2)(X_1+X_2-X_4)}.
\end{equation}

\begin{theorem}
For any $\delta\in\mathbb  C\setminus\{0\}$, define the Painlev\'e deformation map corresponding to the translation $\nu\mapsto\widehat\nu=\nu+2\delta$ by 
$$
L:\quad \left\{\begin{array}{ccl} \widehat X_1 & = & X_1X_4, \\ 
\widehat X_2 & = & X_2X_4, \\
\widehat X_3 & = & X_3X_4-\big(\lambda(\widehat\nu)-\lambda(\nu)\big)Q_\infty(X)\\
                      & = & X_3X_4+\delta(\nu+\delta)(X_1+X_2)(X_1+X_2-X_4),\\
\widehat X_4 & = & X_4^2.
\end{array}\right.
$$
Then, in pencil-adapted coordinates, the map $L$ acts as follows:
\begin{equation}\label{dPE7 map L}
L: (x,y,\nu)\mapsto (\widehat x,\widehat y,\widehat \nu), \quad \widehat x=x+\frac{\delta}{\nu}(x+y),\quad \widehat y=y+\frac{\delta}{\nu}(x+y),\quad \widehat\nu=\nu+2\delta.
\end{equation}
For the latter map,  the factorizations \eqref{L fact 1}, \eqref{L fact 2} are given by
$$
L_1=R_1: (x,y,\nu)\mapsto (x,\widetilde y,\nu+\delta), \quad L_2=R_2:(x,y,\nu)\mapsto (\widetilde x,y, \nu+\delta),
$$
where
\begin{equation}\label{dPE7 L1}
 \widetilde y= y+\frac{\delta}{\nu}(x+y) \quad \Leftrightarrow\quad 
 \frac{\t y+x}{\t y+x -\nu-\delta}=\frac{y+x}{y+x-\nu},
\end{equation}
\begin{equation}\label{dPE7 L2}
\widetilde x= x+\frac{\delta}{\nu}(x+y) \quad \Leftrightarrow\quad
 \frac{\t x+y}{\t x+y -\nu-\delta}=\frac{x+y}{x+y-\nu}.
\end{equation}
\end{theorem}

\paragraph{Relation to the $d$-Painlev\'e equation of the surface type $A_1^{(1)}$.} 

We now compute the 3D Painlev\'e map $\t F=R_1\circ i_1\circ L_1\circ R_2\circ i_2\circ L_2$ in the pencil-adapted coordinates $(x,y,\nu)$. For each fixed $\nu$, the intersection curves $Q_{\lambda(\nu)}\cap P_\mu$ form a pencil through eight points
\begin{equation}
s_i(\nu)=(a_i,-a_i), \quad i=1,\ldots,4, 
\end{equation}
\begin{equation}
s_i(\nu)=\Big(\frac{\nu-1}{2}+a_i,\frac{1+\nu}{2}-a_i\Big), \quad i=5,\ldots,8, 
\end{equation}
which are just the points $S_1,\ldots,S_8$ (which are, recall, independent of $\nu$) expressed in the pencil-adapted coordinates on $Q_{\lambda(\nu)}$. The curve $C_\infty(\nu)$, which is the image of the base curve of the pencil $\{Q_\lambda\}$ in the pencil-adapted coordinates on $Q_{\lambda(\nu)}$, is given by the equation
\begin{equation}
C_\infty(\nu)=\{(x+y)(x+y-\nu)=0\}.
\end{equation}
The map $L$ sends $C_\infty(\nu)$ to $C_\infty(\nu+2\delta)$, while the maps $L_1=R_1$ and $L_2=R_2$ send $C_\infty(\nu)$ to $C_\infty(\nu+\delta)$. We observe that the map $L$ fixes the $(x,y)$ coordinates of the points of the component $\{x+y=0\}$ of $C_\infty(\nu)$, and acts as $(x,y)\mapsto (x+\delta,y+\delta)$ on the component $\{x+y=\nu\}$. This ``shift'' under the map $L$ is, however, only apparent, as this map fixes the curve $\mathcal Q_0\cap\mathcal Q_\infty$ pointwise. Similarly, the map $L_1=R_1$ acts on the second component as $(x,y)\mapsto (x,y+\delta)$, while $L_2=R_2$ acts as $(x,y)\mapsto (x+\delta,y)$. These actions are non-trivial in homogeneous coordinates $X$.

The formulas for the 3D QRT involutions $i_1$, $i_2$ restricted to $Q_{\lambda(\nu)}$  coincide, in the pencil-adapted coordinates, with the original QRT involutions \eqref{dPE7 QRT i1} and \eqref{dPE7 QRT i2}, upon replacing $s_i$ by $s_i(\nu)$:
\begin{eqnarray} \label{dPE7 QRT i1 on Q lambda}
i_1(x,y)=(x,\t y),\quad  \frac{(\t y+x)(x+y)}{(\t y+x-\nu)(x+y-\nu)}=\frac{\prod_{i=1}^{4}(x-a_i)}{\prod_{i=5}^{8}(x-a_i-\frac{\nu-1}{2})}=:\psi_1(x,\nu),
\end{eqnarray}
\begin{eqnarray} \label{dPE7 QRT i2 on Q lambda}
i_2(x,y)=(\t x,y),\quad  \frac{(\t x+y)(x+y)}{(\t x+y-\nu)(x+y-\nu)}=\frac{\prod_{i=1}^{4}(y+a_i)}{\prod_{i=5}^{8}(y+a_i-\frac{1+\nu}{2})}=:\psi_2(y,\nu).
\end{eqnarray}
In the notation of the equations \eqref{dP def eq 1}, \eqref{dP def eq 2}, we have:
\begin{equation}\label{dPE7 aux1}
\frac{(\t x+y_n)(y_n+x)}{(\t x+y_n-\nu_{2n})(y_n+x-\nu_{2n})}=\psi_2(y_n,\nu_{2n}),
\end{equation}
\begin{equation}\label{dPE7 aux2}
\frac{(\t y+x_{n+1})(x_{n+1}+y)}{(\t y+x_{n+1}-\nu_{2n+1})(x_{n+1}+y-\nu_{2n+1})}=\psi_1(x_{n+1},\nu_{2n+1}).
\end{equation}
It remains to express $x,y,\t x,\t y$ in these formulas in terms of $x_n,y_n$. According to \eqref{dP def eq 1}, we have:
$$
L_2:(x_n,y_{n},\nu_{2n-1/2})\mapsto (x,y_n,\nu_{2n})\quad {\rm and}\quad 
R_2:(\t x,y_n,\nu_{2n})\mapsto (x_{n+1},y_n,\nu_{2n+1/2}), 
$$
and with expressions \eqref{dPE7 L2} for the maps $L_2$, $R_2$, we find:
\begin{equation}\label{dPE7 aux 1.1}
\frac{x+y_n}{x+y_n-\nu_{2n}}=\frac{x_n+y_n}{x_n+y_n-\nu_{2n-1/2}},
\end{equation}
\begin{equation}\label{dPE7 aux 1.2}
\frac{\t x+y_n}{\t x+y_n-\nu_{2n}}=\frac{x_{n+1}+y_n}{x_{n+1}+y_n-\nu_{2n+1/2}}.
\end{equation}
Similarly, according to \eqref{dP def eq 2},  we have:
$$
L_1: (x_{n+1},y_n,\nu_{2n+1/2})\mapsto (x_{n+1},y,\nu_{2n+1}) \quad {\rm and}\quad 
R_1: (x_{n+1},\t y,\nu_{2n+1})\mapsto (x_{n+1},y_{n+1},\nu_{2n+3/2}),
$$
and with expressions \eqref{dPE7 L1} for the maps $L_1$, $R_1$, we find:
\begin{equation}\label{dPE7 aux 2.1}
\frac{y+x_{n+1}}{y+x_{n+1}-\nu_{2n+1}}=\frac{x_{n+1}+y_{n}}{x_{n+1}+y_{n}-\nu_{2n+1/2}},
\end{equation}
\begin{equation}\label{dPE7 aux 2.2}
\frac{x_{n+1}+\t y}{x_{n+1}+\t y-\nu_{2n+1}}=\frac{x_{n+1}+y_{n+1}}{x_{n+1}+y_{n+1}-\nu_{2n+3/2}}.
\end{equation}
Combining equations \eqref{dPE7 aux1}, \eqref{dPE7 aux2} with \eqref{dPE7 aux 1.1}--\eqref{dPE7 aux 2.2} results in the following non-autonomous system: 
\begin{equation}
\frac{(x_{n+1}+y_n)(x_n+y_n)}{(x_{n+1}+y_n-\nu_{2n+1/2})(x_n+y_n-\nu_{2n-1/2})}=\psi_2(y_n,\nu_{2n}),
\end{equation}
\begin{equation}
\frac{(x_{n+1}+y_{n+1})(x_{n+1}+y_{n})}{(x_{n+1}+y_{n+1}-\nu_{2n+3/2})(x_{n+1}+y_{n}-\nu_{2n+1/2})}=\psi_1(x_{n+1},\nu_{2n+1}).
\end{equation}
This is nothing but the $d$-Painlev\'e equation of the surface type $A_1^{(1)}$, as given in \cite{KNY}.
\smallskip

{\bf Remark.} The symmetric case can be characterized by $\psi_1(x,\nu)=\psi_2(x,\nu)$. In this case the latter equations become two instances of 
\begin{equation}
\frac{(u_{n+1}+u_n)(u_n+u_{n-1})}{(u_{n+1}+u_n-\nu_{n+1/2})(u_n+u_{n-1}-\nu_{n-1/2})}=\psi_1(u_n,\nu_{n}),
\end{equation}
if we set $u_{2n-1}=x_n$, $u_{2n}:=y_n$.

%%%%%%%%%%%%%%%%%%%%%%%%%%%%%%%
%%%%%%%%%%%%%%%%%%%%%%%%%%%%%%%
\section{From a pencil of type (vi) to the \textit{d}-Painlev\'e equation of the surface type $D_4^{(1)}$}
\label{sect dPD4}
%%%%%%%%%%%%%%%%%%%%%%%%%%%%%%%
%%%%%%%%%%%%%%%%%%%%%%%%%%%%%%%

By a simple limiting procedure, the results of the previous section lead to similar results for the \textit{d}-Painlev\'e equation of the surface type $D_4^{(1)}$. We refrain from giving complete details here, and restrict ourselves only to the main results.

\paragraph{2D QRT map.}
We consider the QRT map corresponding to the pencil of biquadratic curves $\{C_\mu\}$ through eight points 
\begin{equation}
s_i=(a_i,-a_i), \;\;  s_{i+4}=(a_i+\epsilon,-a_i+\epsilon),\;\;i=1,\ldots,4,
\end{equation}
where the points $s_5,\ldots,s_8$ are infinitely near to  $s_1,\ldots,s_4$, respectively. 
This pencil contains a reducible curve: 
\begin{equation}
C_\infty=\{(x+y)^2=0\}.
\end{equation}
The vertical involution $i_1$ and the horizontal involution $i_2$ for this pencil can be described by the following equations:
 \begin{equation} \label{dPD4 QRT i1}
 i_1(x,y)=(x,\t y),\quad  \frac{1}{\t y+x}+\frac{1}{x+y}=\frac{1}{2}\sum_{i=1}^4\frac{1}{x-a_i},
 \end{equation}
 \begin{equation} \label{dPD4 QRT i2}
i_2(x,y)=(\t x,y),\quad   \frac{1}{\t x+y}+\frac{1}{x+y}=\frac{1}{2}\sum_{i=1}^4\frac{1}{y+a_i}.
 \end{equation}

%%%%%%%%%%%%%%%%%%%%%%%%%%%%%%%%%
\begin{figure}[!ht]
\begin{center}
\begin{subfigure}[b]{0.3\textwidth}
         \centering
         \includegraphics[width=0.8\textwidth]{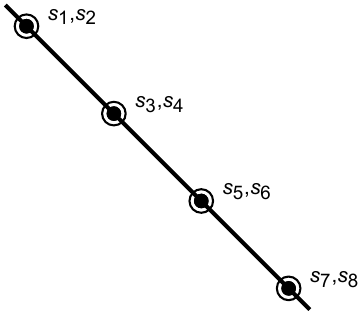}
         \caption{}
     \end{subfigure}
     \hspace{2cm}
     \begin{subfigure}[b]{0.3\textwidth}
         \centering
         \includegraphics[width=\textwidth]{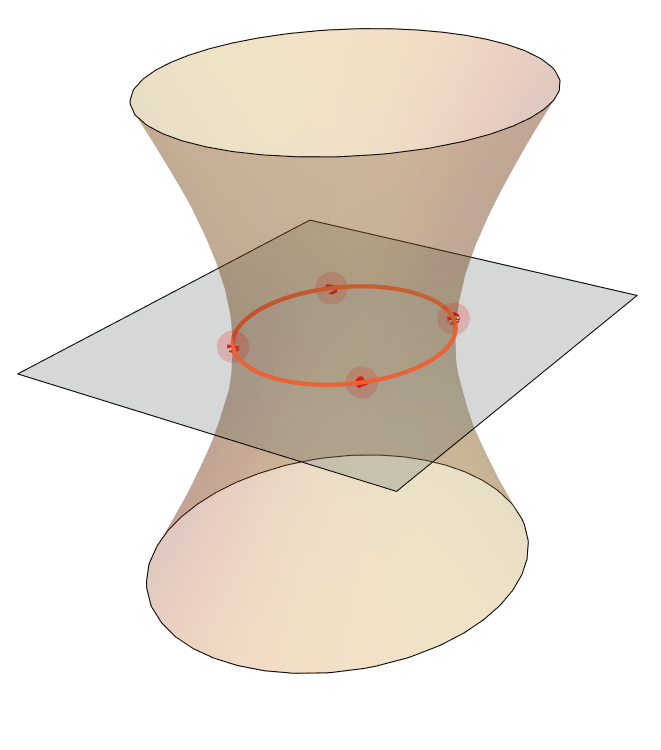}
         \caption{}
     \end{subfigure}
\end{center}
\caption{(a) Base set of the surface type $D_4^{(1)}$: four double points on a double (1,1)-curve in $\bbP^1\times\bbP^1$. (b) Pencil of quadrics touching along a conic}
\label{fig dPD4}
\end{figure}
%%%%%%%%%%%%%%%%%%%%%%%%%%%%%%%%%

\paragraph{3D Painlev\'e map.} We consider the pencil of quadrics $\{P_\mu\}$ in $\bbP^3$ obtained as the Segre lift of the pencil of curves $\{C_\mu\}$. The pencil $\{Q_\lambda\}$ is spanned by $Q_0=\{X_1X_2-X_3X_4=0\}$ and
$Q_\infty=P_{\infty}=\{(X_1+X_2)^2=0\}$:
\begin{equation}\label{dPD4 pencil}
Q_\lambda=\Big\{X_1X_2-X_3X_4-\lambda(X_1+X_2)^2=0\Big\}.
\end{equation}
The base set of the pencil $Q_\lambda$ is the double conic $\{X_1X_2-X_3X_4=0,\; X_1+X_2=0\}$. This is a pencil of type (vi). 
The matrix $M_\lambda$ of the quadratic form $Q_\lambda$ is:
\begin{equation}
M_\lambda=\begin{pmatrix}
-2\lambda & 1-2\lambda & 0  & 0\\
1-2\lambda & -2\lambda & 0 & 0 \\
0 & 0 & 0 & -1 \\
0 & 0 & -1 & 0 
\end{pmatrix}.
\end{equation}
The characteristic polynomial of the pencil $\{Q_\lambda\}$ is: $\Delta(\lambda)=\det(M_\lambda)=-1+4\lambda$, the same as in Section \ref{sect dPE7}. The normalizing transformation of $Q_{\lambda(\nu)}$ to the canonical form $Q_0$ reads: 
\begin{equation}\label{dPD4 pencil norm}
\begin{bmatrix} X_1\\X_2\\X_3\\X_4\end{bmatrix} = A_\nu \! \begin{bmatrix} Y_1 \\ Y_2 \\ Y_3 \\ Y_4\end{bmatrix}, 
\end{equation}
where
\begin{equation}
A_\nu=\begin{pmatrix} \frac{1}{2\nu}(1+\nu) & \frac{1}{2\nu}(1-\nu) & 0 &  0 \\[0.15cm] 
\frac{1}{2\nu}(1-\nu) & \frac{1}{2\nu}(1+\nu) & 0 & 0 \\[0.15cm]  
0 & 0 & 1 & 0 \\[0.15cm] 
0 & 0 & 0 & 1 \end{pmatrix}.                               
 \end{equation}
A parametrization of the quadric $Q_{\lambda(\nu)}$ is given by:
\begin{equation}\label{dPD4 x to X}
\begin{bmatrix} X_1 \\ X_2 \\ X_3 \\ X_4 \end{bmatrix}=A_\nu \begin{bmatrix} x \\ y \\ xy \\ 1 \end{bmatrix}
=\begin{bmatrix} \frac{1}{2\nu}((1+\nu)x+(1-\nu)y) \\[0.15cm] 
\frac{1}{2\nu}((1-\nu)x+(1+\nu)y) \\[0.15cm] 
xy \\[0.15cm] 1  \end{bmatrix}=:\phi_\nu(x,y).
\end{equation}
The pencil-adapted coordinates $(x,y,\nu)$ on (the double cover of) $\bbP^3$ are:
\begin{equation}\label{dPD4 X to x}
x = \frac{(1+\nu)X_1-(1-\nu)X_2}{2X_4}, \quad y = \frac{(1+\nu)X_2-(1-\nu)X_1}{2X_4},
\end{equation}
which have to be supplemented with
\begin{equation}\label{dPD4 X to lambda}
\lambda=\frac{1-\nu^2}{4}=\frac{X_1X_2-X_3X_4}{(X_1+X_2)^2}.
\end{equation}

\begin{theorem}
For any $\delta\in\mathbb  C\setminus\{0\}$, define the Painlev\'e deformation map corresponding to the translation $\nu\mapsto\widehat\nu=\nu+2\delta$ by 
$$
L:\quad \left\{\begin{array}{ccl} \widehat X_1 & = & X_1X_4, \\ 
\widehat X_2 & = & X_2X_4, \\
\widehat X_3 & = & X_3X_4-\big(\lambda(\widehat\nu)-\lambda(\nu)\big)Q_\infty(X)\\
                      & = & X_3X_4+\delta(\nu+\delta)(X_1+X_2)^2,\\
\widehat X_4 & = & X_4^2.
\end{array}\right.
$$
Then, in pencil-adapted coordinates, the map $L$ acts as follows:
\begin{equation}\label{dPD4 map L}
L: (x,y,\nu)\mapsto (\widehat x,\widehat y,\widehat \nu), \quad \widehat x=x+\frac{\delta}{\nu}(x+y),\quad \widehat y=y+\frac{\delta}{\nu}(x+y),\quad \widehat\nu=\nu+2\delta.
\end{equation}
For the latter map, the factorizations \eqref{L fact 1}, \eqref{L fact 2} are given by
$$
L_1=R_1: (x,y,\nu)\mapsto (x,\widetilde y,\nu+\delta), \quad L_2=R_2:(x,y,\nu)\mapsto (\widetilde x,y, \nu+\delta),
$$
where
\begin{equation}\label{dPD4 L1}
 \widetilde y= y+\frac{\delta}{\nu}(x+y) \quad \Leftrightarrow\quad 
 \frac{\nu+\delta}{\t y+x}=\frac{\nu}{y+x},
\end{equation}
\begin{equation}\label{dPD4 L2}
\widetilde x= x+\frac{\delta}{\nu}(x+y) \quad \Leftrightarrow\quad
 \frac{\nu+\delta}{\t x+y }=\frac{\nu}{x+y}.
\end{equation}
\end{theorem}

Computing the 3D Painlev\'e map $\t F=R_1\circ i_1\circ L_1\circ R_2\circ i_2\circ L_2$ in the pencil-adapted coordinates $(x,y,\nu)$, we come to the following non-autonomous system: 
\begin{equation}
\frac{\nu_{2n+1/2}}{x_{n+1}+y_n}+\frac{\nu_{2n-1/2}}{x_n+y_n}=\frac{\nu_{2n}}{2}\sum_{i=1}^4\frac{1}{y_n+a_i},
\end{equation}
\begin{equation}
\frac{\nu_{2n+3/2}}{x_{n+1}+y_{n+1}}+\frac{\nu_{2n+1/2}}{x_{n+1}+y_{n}}=\frac{\nu_{2n+1}}{2}\sum_{i=1}^4\frac{1} {x_{n+1}-a_i}.
\end{equation}
This can be considered as a $d$-Painlev\'e equation of the surface type $D_4^{(1)}$ (in a realization different from that in \cite{KNY}).
\smallskip

The symmetric case is characterised by $a_{2i}=-a_{2i-1}$, $i=1,2$. In this case the latter equations become two instances of 
\begin{equation}
\frac{\nu_{n+1/2}}{u_{n+1}+u_n}+ \frac{\nu_{n-1/2}}{u_n+u_{n-1}}=\nu_n\Big(\frac{u_n}{u_n^2-a_1^2}+\frac{u_n}{u_n^2-a_3^2}\Big),
\end{equation}
if we set $u_{2n-1}=x_n$, $u_{2n}:=y_n$.

%%%%%%%%%%%%%%%%%%%%%%%%%%%%%%%
%%%%%%%%%%%%%%%%%%%%%%%%%%%%%%%
\section{From a pencil of type (iv) to the \textit{q}-Painlev\'e equation of the surface type $A_1^{(1)}$}
\label{sect qPE7}
%%%%%%%%%%%%%%%%%%%%%%%%%%%%%%%
%%%%%%%%%%%%%%%%%%%%%%%%%%%%%%%

\begin{figure}[!ht]
\begin{center}
\begin{subfigure}[b]{0.3\textwidth}
         \centering
         \includegraphics[width=0.8\textwidth]{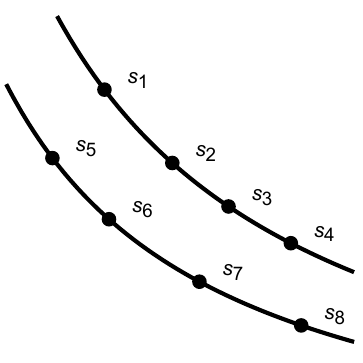}
         \caption{}
     \end{subfigure}
     \hspace{2cm}
     \begin{subfigure}[b]{0.3\textwidth}
         \centering
         \includegraphics[width=\textwidth]{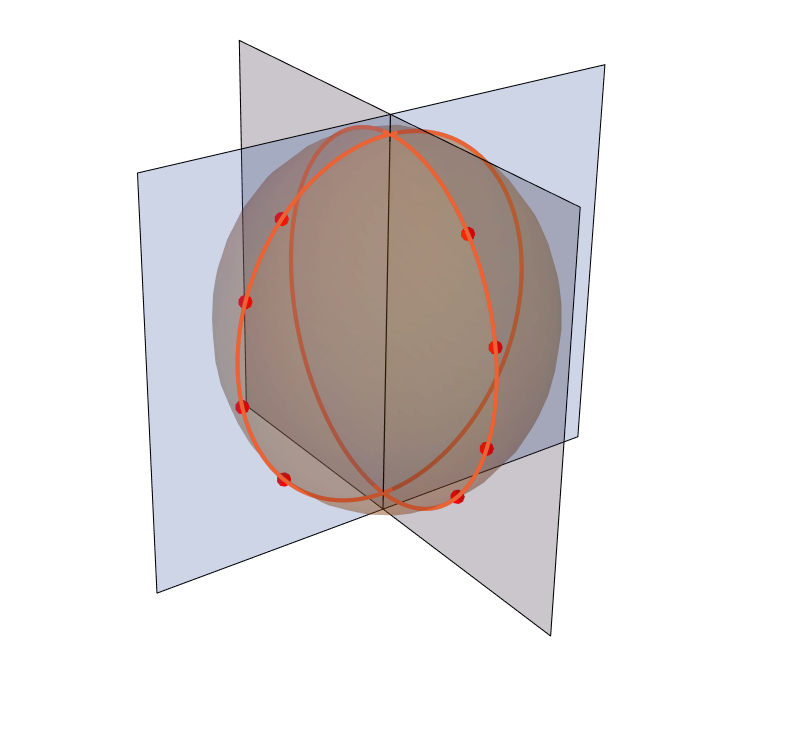}
         \caption{}
     \end{subfigure}
\end{center}
\caption{(a) Base set of the surface type $A_1^{(1)}$: two quadruples of points on two (1,1)-curves (hyperbolas) in $\bbP^1\times\bbP^1$ intersecting at two points $(\infty,0)$ and $(0,\infty)$. (b) Pencil of quadrics through two non-coplanar conics intersecting at two points}
\label{fig qPE7}
\end{figure}

\paragraph{2D QRT map.}
Consider the QRT map corresponding to the pencil of biquadratic curves through eight points 
\begin{eqnarray}
&& s_i=(a_i,b_i)=\big(\ka c_i,\ka c_i^{-1}\big), \quad i=1,\ldots,4,\\
&& s_i=(a_i,b_i)=\big(c_i,c_i^{-1}\big), \quad i=5,\ldots,8,
\end{eqnarray}
with $\kappa \neq 0,1$.
These eight points support a pencil of biquadratic curves if they satisfy the condition 
\begin{equation}\label{qPE7 cond}
   \frac{\prod_{i=1}^4 c_i}{\prod_{i=5}^8 c_i}=1\quad\Leftrightarrow\quad \frac{\prod_{i=1}^4 a_i}{\prod_{i=5}^8 a_i}=\ka^{4} \quad\Leftrightarrow\quad \frac{\prod_{i=1}^4 b_i}{\prod_{i=5}^8 b_i}=\ka^{4}.
\end{equation}
They are symmetric with respect to $\sigma(x,y)=(y,x)$ if $c_{2i}=c_{2i-1}^{-1}$, $i=1,\ldots,4$.
See Fig. \ref{fig qPE7} (a).
%%%%%%%%%%%%%%%%%%%%%%%%%%%%%%%%%

%%%%%%%%%%%%%%%%%%%%%%%%%%%%%%%%%
This pencil contains a reducible curve consisting of two (1,1)-curves:
\begin{equation}
C_\infty=\big\{(xy-1)(xy-\ka^2)=0\big\}.
\end{equation}

The vertical involution $i_1$ can be described by the following equation:
 \begin{equation} \label{qPE7 QRT i1}
 i_1(x,y)=(x,\t y),\quad  \frac{(x\t y-\ka^2)(xy-\ka^2)}{(x\t y-1)(xy-1)}=\frac{\prod_{i=1}^{4}(x-\ka c_i)}{\prod_{i=5}^{8}(x-c_i)}.
 \end{equation}
Similarly, the horizontal involution $i_2$ can be described by the following equation:
 \begin{equation} \label{qPE7 QRT i2}
i_2(x,y)=(\t x,y),\quad   \frac{(\t x y-\ka^2)(xy-\ka^2)}{(\t xy-1)(xy-1)}=\frac{\prod_{i=1}^{4}(y- \ka c_i^{-1})}{\prod_{i=5}^{8}(y-c_i^{-1})}.
 \end{equation}
The QRT map $F$ is the composition of these two involutions, $F=i_1\circ i_2$. The singularity confinement structure of the QRT involutions is as in \eqref{sing1}. In the symmetric case we have $F=f^2$, with  $f=i_1\circ \sigma=\sigma\circ i_2$ being the QRT root.

\paragraph{3D Painlev\'e map.} As usual, we identify $\bbP^1\times\bbP^1$ with the quadric 
$$
Q_0=\{X_1X_2-X_3X_4=0\}\subset\bbP^3,
$$ 
via $[X_1:X_2:X_3:X_4]=[x:y:xy:1]$.  The points $s_i$ are lifted to
\begin{equation}\label{qPE7 S}
S_i=[a_i:b_i:a_ib_i:1]=\left\{\begin{array}{l} \big[\ka c_i:\ka c_i^{-1}:\ka^2:1\big], \;i=1,\ldots,4, \\[0.25cm] 
\big[c_i:c_i^{-1}:1:1\big], \;i=5,\ldots,8.\end{array}\right.
\end{equation}
We declare $Q_\lambda$ to be spanned by $Q_0$ and $Q_\infty=P_{\infty}=(X_3-\ka X_4) (\ka X_3- X_4)$:
\begin{equation}\label{qPE7 pencil}
Q_\lambda=\Big\{X_1X_2-X_3X_4-\lambda(X_3-\ka^2 X_4) (X_3-X_4)=0\Big\}.
\end{equation}
The base set of the pencil $Q_\lambda$ consists of two conics, $\{X_1X_2-X_3X_4=0,\; X_3-\ka^2 X_4=0\}$ and  $\{X_1X_2-X_3X_4=0,\; X_3- X_4=0\}$, which intersect at two points $[0:1:0:0]$ and $[1:0:0:0]$. This is a pencil of type (iv).

The matrix $M_\lambda$ of the quadratic form $Q_\lambda$:
\begin{equation}
M_\lambda=\begin{pmatrix}
0 & 1 & 0  & 0 \\
1 & 0 & 0 & 0 \\
0 & 0 & -2 \lambda & -1-(1+\ka^2)\lambda \\
0 & 0 & -1-(1+\ka^2)\lambda & -2\ka^2 \lambda 
\end{pmatrix}
\end{equation}
The characteristic polynomial of the pencil $\{Q_\lambda\}$ is: 
$$
\Delta(\lambda)=\det(M_\lambda)=\big(1+(1+\ka^2)\lambda\big)^2-4\ka^2\lambda^2=\big(1+(1+\ka)^2\lambda\big)\big(1+(1-\ka)^2\lambda\big),
$$ 
so that ${\rm Sing}(Q_\lambda)=\{-(1+\ka)^{-2},-(1-\ka)^{-2},\infty\}$. This polynomial is not a complete square, and we have to uniformize $\sqrt{\Delta(\lambda)}$. The uniformizing variable is $\nu\in\bbC$. However, in the present situation it will be convenient to use $w=e^\nu$ instead, with $w\in \bbC\setminus \{0\}$. The shift $\nu\mapsto\nu+\delta$ will be replaced by $w\mapsto qw$ with $q=e^\delta$. We set
\begin{equation}\label{qPE7 lambda unif}
	\lambda =\lambda(w)= \dfrac{(\ka-w)(1-\ka w)}{(1-\ka^2)^2 w}.
\end{equation} 
Then $\Delta(\lambda)$ becomes a square:
$$
\Delta(\lambda)=\dfrac{\ka^2(1-w^2)^2}{w^2(1-\ka^2)^2}\quad \Rightarrow\quad \sqrt{\Delta(\lambda)}=\dfrac{\ka(1-w^2)}{w(1-\ka^2)}.
$$
Observe that $\lambda(w)=\lambda(w^{-1})$, while $\sqrt{\Delta(\lambda)}$ changes its sign under $w \mapsto w^{-1}$. This gives us a double cover of the original pencil branched at $\lambda=(1+\ka)^{-2}$ (corresponding to $w=1$), and at $\lambda=(1-\ka)^{-2}$ (corresponding to $w=-1$). The point $\lambda=\infty$ (corresponding to $w=0,\infty$) is not a branch point. The normalizing transformation of $Q_\lambda(X)$ to the canonical form $Q_0(Y)=Y_1Y_2-Y_3Y_4$ is achieved by the transformation
\begin{equation}\label{hyp qPE8 pencil norm}
\begin{bmatrix} X_1\\X_2\\X_3\\X_4\end{bmatrix} = A_w \!\begin{bmatrix} Y_1 \\ Y_2 \\ Y_3 \\ Y_4\end{bmatrix}, 
\end{equation}
where one can take
\begin{equation}
A_w=\begin{pmatrix} 
1 & 0  &  0 & 0 \\[0.25cm] 
0 & 1  & 0 & 0 \\[0.25cm]
0&0 & \dfrac{1-\ka w}{1-w^2} & \dfrac{w(\ka-w)}{1-w^2}\\[0.2cm]
0 & 0 &  \dfrac{\ka-w}{\ka(1-w^2)} &  \dfrac{w(1-\ka w)}{\ka(1-w^2)}
 	 \end{pmatrix}.                          
 \end{equation}
 Indeed, one immediately verifies that 
 $$
 A_w^{\rm T} M_{\lambda(w)} A_w=M_0.
 $$
Now, we are in the position to derive a parametrization of the quadric $Q_\lambda$:
\begin{equation}\label{hyp qPE7 x to X}
\begin{bmatrix} X_1 \\[0.2cm] X_2 \\[0.2cm] X_3 \\[0.2cm] X_4 \end{bmatrix} =A_w
 	\begin{bmatrix} x \\[0.2cm] y \\[0.2cm] xy \\[0.2cm] 1 \end{bmatrix} =: \phi_w(x,y).
\end{equation}
Observe that this parametrization is neither valid for $w=0$ nor for $w=\infty$.
The pencil-adapted coordinates $(x,y,w)$ on (the double cover of) $\bbP^3$ are:
\begin{equation}\label{qPE7 X to x}
	x = \dfrac{(1-\ka w)X_3-\ka(\ka-w)X_4}{(1-\ka^2)X_2}=\dfrac{w(1-\ka^2)X_1}{\ka(1-\ka w)X_4-(\ka-w)X_3},
\end{equation}
\begin{equation}\label{qPE7 X to y}
	y = \dfrac{(1-\ka w)X_3-\ka(\ka-w)X_4}{(1-\ka^2)X_1}= \dfrac{w(1-\ka^2)X_2}{\ka(1-\ka w)X_4-(\ka-w)X_4},
\end{equation}
which have to be supplemented with
\begin{equation}\label{qPE7 X to lambda}
\lambda = \dfrac{(\ka-w)(1-\ka w)}{(\ka^2-1)^2 w} = \dfrac{X_1 X_2-X_3X_4}{(X_3-\ka^2 X_4)(X_3-X_4)}.
\end{equation}

\begin{theorem}
For any $q\neq \pm 1$, define the Painlev\'e deformation map corresponding to the translation $w\mapsto \widehat w=q^2 w$ by 
\begin{equation}\label{qPE7 map L in X}
L: \quad\left\{ \begin{array}{l} \widehat X_1 = X_1X_2+\big(\lambda(\widehat w)-\lambda(w)\big)(X_3-X_4)(X_3-\ka^2 X_4),\\
\widehat X_2=X_2^2,\\
\widehat X_3=X_2X_3,\\
\widehat X_4=X_2X_4,
\end{array}\right.
\end{equation}
where $\lambda=\lambda(w)$ is given by \eqref{qPE7 lambda unif}. Then, in pencil-adapted coordinates, the map $L$ acts as follows:
\begin{equation}\label{qPE7 map L}
L: \quad \widehat x=\frac{q^2w^2-1}{w^2-1}x-\frac{(q^2-1)w^2}{w^2-1}y^{-1},\quad 
\widehat y^{\ -1}=\frac{q^2w^2-1}{q^2(w^2-1)}y^{-1}-\frac{(q^2-1)}{q^2(w^2-1)}x, \quad \widehat w=q^2 w.
\end{equation}
For the latter map,  the factorizations \eqref{L fact 1}, \eqref{L fact 2} are given by
$$
L_1=R_1:  (x,y,w)\mapsto (x,\t y,qw), \quad L_2=R_2: (x,y,w)\mapsto (\t x,y,qw)
$$
where
\begin{equation}\label{qPE7 L1}
\t y^{\ -1}=\frac{q^2w^2-1}{q^2(w^2-1)}y^{-1}-\frac{(q^2-1)}{q^2(w^2-1)}x\quad\Leftrightarrow\quad 
\frac{\t{y} x-q^2w^2}{\t{y} x-1}=q^2\,\frac{y x-w^2}{y x-1},
\end{equation}
and
\begin{equation}\label{qPE7 L2}
\t x=\frac{q^2w^2-1}{w^2-1}x-\frac{(q^2-1)w^2}{w^2-1}y^{-1} \quad \Leftrightarrow\quad 
\frac{\t{x} y-q^2w^2}{\t{x} y-1}=\frac{x y-w^2}{xy-1}.
\end{equation}
\end{theorem}

\paragraph{Relation to the $q$-Painlev\'e equation of the surface type $A_1^{(1)}$.} 

We now compute the 3D Painlev\'e map $\t F=R_1\circ i_1\circ L_1\circ R_2\circ i_2\circ L_2$ in the pencil-adapted coordinates $(x,y,w)$. 
For each fixed $w$, the intersection curves $Q_{\lambda(w)}\cap P_\mu$ form a pencil through eight points
\begin{equation}
s_i(w)=\big(w c_i,w c_i^{-1}\big), \quad i=1,\ldots,4, 
\end{equation}
\begin{equation}
s_i(w)=\big(c_i,c_i^{-1}\big), \quad i=5,\ldots,8, 
\end{equation}
which are just the points $S_1,\ldots,S_8$ expressed in the pencil-adapted coordinates on $Q_{\lambda(w)}$.
The formulas for the 3D QRT involutions $i_1$, $i_2$ restricted to $Q_{\lambda(w)}$  coincide, in the pencil-adapted coordinates, with the original QRT involutions \eqref{qPE7 QRT i1} and \eqref{qPE7 QRT i2}, upon replacing $\ka$ by $w$, and $s_i$ by $s_i(w)$:
 \begin{equation} \label{qPE7 QRT i1 on Q lambda}
 i_1(x,y)=(x,\t y),\quad  \frac{(x\t y-w^2)(xy-w^2)}{(x\t y-1)(xy-1)}=\frac{\prod_{i=1}^{4}(x-w c_i)}{\prod_{i=5}^{8}(x-c_i)},
 \end{equation}
 \begin{equation} \label{qPE7 QRT i2 on Q lambda}
i_2(x,y)=(\t x,y),\quad   \frac{(\t x y-w^2)(xy-w^2)}{(\t xy-1)(xy-1)}=\frac{\prod_{i=1}^{4}(y- w c_i^{-1})}{\prod_{i=5}^{8}(y-c_i^{-1})}.
 \end{equation}
 In the notation of the equations \eqref{dP def eq 1}, \eqref{dP def eq 2}, the latter two equations read:
\begin{equation}\label{qPE7 aux1} 
\frac{(\t x y_n-w_{2n}^2)(x y_n-w_{2n}^2)}{(\t x y_n-1)(xy_n-1)}= \frac{\prod_{i=1}^{4}(y_n-w_{2n} c_i^{-1})}{\prod_{i=5}^{8}(y_n-c_i^{-1})},
\end{equation}
\begin{equation}\label{qPE7 aux2}
\frac{(\t y x_{n+1}-w_{2n+1}^2)(y  x_{n+1}-w_{2n+1}^2)}{(\t yx_{n+1}-1)(y x_{n+1}-1)}=\frac{\prod_{i=1}^{4}(x_{n+1}-w_{2n+1} c_i)}{\prod_{i=5}^{8}(x_{n+1}-\ c_i)},
\end{equation}
where
\begin{equation}
w_{2n-1/2}=q^{-1}w_{2n}, \quad w_{2n+1/2}=qw_{2n}.
\end{equation}
According to \eqref{dP def eq 1}, we have:
$$
L_2:(x_n,y_{n},w_{2n-1/2})\mapsto (x,y_n,w_{2n})\quad {\rm and}\quad 
R_2:(\t x,y_n,w_{2n})\mapsto (x_{n+1},y_n,w_{2n+1/2}). 
$$
With expressions \eqref{qPE7 L2} for the maps $L_2$, $R_2$, we find:
\begin{equation}\label{qPE7 aux 1.1}
\frac{xy_n-w_{2n}^2 }{xy_n-1}=\frac{x_n y_n-w_{2n}w_{2n-1}}{x_ny_n-1},
\end{equation}
\begin{equation}\label{qPE7 aux 1.2}
\frac{\t x y_n-w_{2n}^2}{\t x y_n-1}=\frac{x_{n+1} y_n-w_{2n+1}w_{2n}}{x_{n+1} y_n-1}.
\end{equation}
Similarly, according to \eqref{dP def eq 2}, we have:
$$
L_1: (x_{n+1},y_n,w_{2n+1/2})\mapsto (x_{n+1},y,w_{2n+1}) \quad {\rm and}\quad 
R_1: (x_{n+1},\t y,w_{2n+1})\mapsto (x_{n+1},y_{n+1},w_{2n+3/2}),
$$
and with expressions \eqref{qPE7 L1} for the maps $L_1$, $R_1$, we find:
\begin{equation}\label{qPE7 aux 2.1}
q^{-2} \frac{ x_{n+1}y-w_{2n+1}^2}{x_{n+1}y-1}= \frac{x_{n+1}y_n-w_{2n+1}w_{2n}}{x_{n+1}y_n-1},
\end{equation}
\begin{equation}\label{qPE7 aux 2.2}
q^2 \frac{\t yx_{n+1}-w_{2n+1}^2}{\t yx_{n+1}-1}=\frac{x_{n+1}y_{n+1}-w_{2n+2}w_{2n+1}}{x_{n+1}y_{n+1}-1}.
\end{equation}
Combining equations \eqref{qPE7 aux1}, \eqref{qPE7 aux2} with \eqref{qPE7 aux 1.1}--\eqref{qPE7 aux 2.2} results in the following non-autonomous system: 
\begin{equation}
\frac{(x_{n+1}y_n-w_{2n+1}w_{2n})(x_n y_n-w_{2n}w_{2n-1})}{(x_{n+1} y_n-1)(x_ny_n-1)}= 
\frac{\prod_{i=1}^{4}(y_n-w_{2n} c_i^{-1})}{\prod_{i=5}^{8}(y_n- c_i^{-1})},
\end{equation}
\begin{equation}
 \frac{(y_{n+1} x_{n+1}-w_{2n+2}w_{2n+1})(y_n x_{n+1}-w_{2n+1}w_{2n})} {(y_{n+1} x_{n+1}-1)(y_n x_{n+1}-1)}=
\frac{\prod_{i=1}^{4}(x_{n+1}-w_{2n+1} c_i)}{\prod_{i=5}^{8}(x_{n+1}-c_i)}.
 \end{equation}
This is the \textit{q}-Painlev\'e equation of the surface type $A_1^{(1)}$ , as given in \cite{KNY}. In the symmetric case, if $c_{2i}=c_{2i-1}^{-1}$, $i=1,\ldots,4$, these equations become two instances of 
\begin{equation}
\frac{(u_{n+1} u_n-w_{n+1}w_n)( u_nu_{n-1}-w_{n}w_{n-1})}{(u_{n+1}u_n-1)(u_nu_{n-1}-1)}=
\frac{\prod_{i=1}^{4}(u_n-w_{n} c_i)}{\prod_{i=5}^{8}(u_{n}- c_i)}.
\end{equation}

%%%%%%%%%%%%%%%%%%%%%%%%%
%%%%%%%%%%%%%%%%%%%%%%%%%
\section{From a pencil of type (iii) to the \textit{d}-Painlev\'e equation of the surface type $A_0^{(1)}$}
\label{sect dPE8}
%%%%%%%%%%%%%%%%%%%%%%%%%%
%%%%%%%%%%%%%%%%%%%%%%%%%%

\paragraph{2D QRT map.} 

\begin{figure}[!ht]
\begin{center}
\begin{subfigure}[b]{0.3\textwidth}
         \centering
         \includegraphics[width=0.8\textwidth]{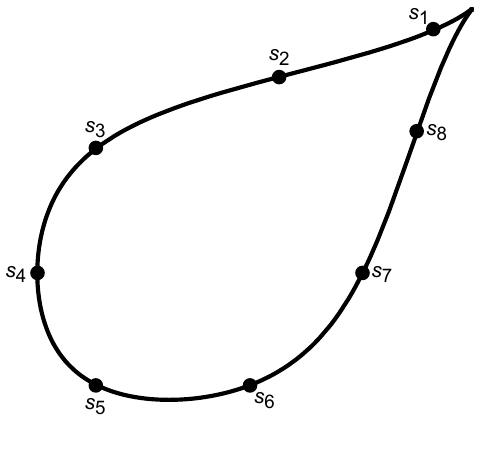}
         \caption{}
     \end{subfigure}
     \hspace{2cm}
     \begin{subfigure}[b]{0.3\textwidth}
         \centering
         \includegraphics[width=\textwidth]{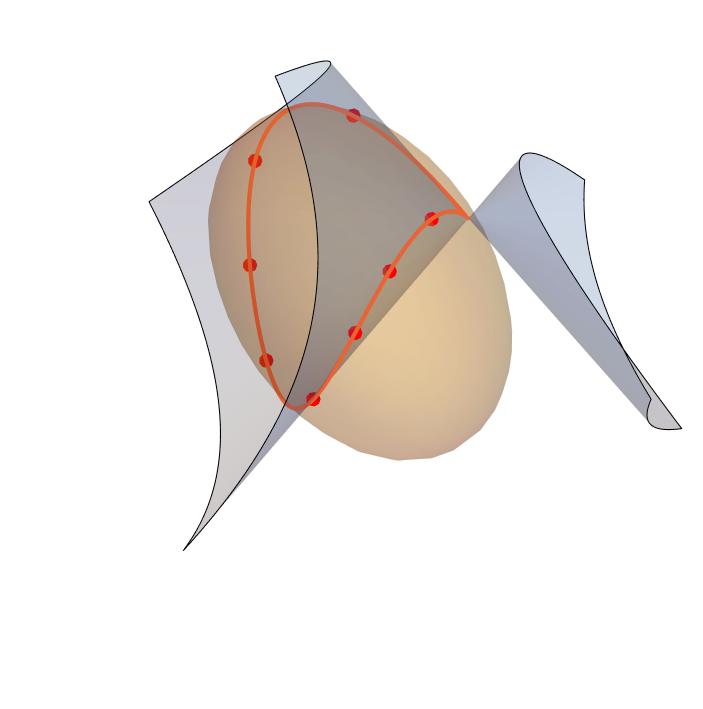}
         \caption{}
     \end{subfigure}
\end{center}
\caption{(a) Base set of the surface type $A_0^{(1)}$: eight points on a cuspidal (2,2)-curve in $\bbP^1\times\bbP^1$. (b) Pencil of quadrics through cuspidal spatial quartic in $\bbP^3$}
\label{fig dPE8}
\end{figure}

We consider the QRT map corresponding to the pencil of biquadratic curves through eight points $s_i=(a_i,b_i)$, $i=1,\ldots,8$, where
$$
a_i=z_i(z_i-\ka_1), \quad b_i=z_i(z_i-\ka_2).
$$
These eight points support a pencil of biquadratic curves if they satisfy the condition
$$
\sum_{i=1}^8 z_i=2(\ka_1+\ka_2).
$$
They belong to the curve with the equation
$$
(x-y)^2=(\ka_2-\ka_1)(\ka_2 x-\ka_1 y).
$$
This is a biquadratic curve in $\bbP^1\times\bbP^1$ with a cusp point at $(\infty,\infty)$, see Fig. \ref{fig dPE8} (a).

%%%%%%%%%%%%%%%%%%%%%%%%%%%%%%%%%
%%%%%%%%%%%%%%%%%%%%%%%%%%%%%%%%%

The vertical involution $i_1$ can be described by the following equation:
 \begin{align} \label{dPE8 QRT i1}
&  i_1(x,y)=(x,\t y),\nonumber\\ 
& \frac{\big(\t y-\xi(\xi-\ka_2)\big)\big(y-\xi(\xi-\ka_2)\big)}{\big(\t y-(\xi-\ka_1)(\xi-\ka_1+\ka_2)\big)
 \big(y-(\xi-\ka_1)(\xi-\ka_1+\ka_2)\big)}=\frac{U(\xi)}{U(\ka_1-\xi)}, \quad x=\xi(\xi-\ka_1).
 \end{align}
 Here we use the abbreviation 
 \begin{equation}\label{dPE8 U}
 U(z)=\prod_{i=1}^8 (z-z_i).
 \end{equation}
Formula \eqref{dPE8 QRT i1} is understood as follows. Written as a polynomial in $\xi$, it is anti-symmetric with respect to $\xi\leftrightarrow \ka_1-\xi$. Upon division by $\xi-2\ka_1$, the resulting polynomial is symmetric and therefore it can be actually expressed as a polynomial in $x=\xi(\xi-\ka_1)$. This defines $i_1$ as a birational involution (its symmetry w.r.t. $y\leftrightarrow \t y$ is obvious).

Similarly, the horizontal involution $i_2$ can be described by the following equation:
 \begin{align} \label{dPE8 QRT i2}
& i_2(x,y)=(\t x,y),\nonumber\\  
& \frac{\big(\t x-\eta(\eta-\ka_1)\big)\big(x-\eta(\eta-\ka_1)\big)}{\big(\t x-(\eta-\ka_2)(\eta-\ka_2+\ka_1)\big)\big(x-(\eta-\ka_2)(\eta-\ka_2+\ka_1)\big)}=\frac{U(\eta)}{U(\ka_2-\eta)}, \quad y=\eta(\eta-\ka_2).
 \end{align}
The QRT map $F$ is the composition of these two involutions, $F=i_1\circ i_2$. 
The eight points $s_1,\ldots,s_8$ in $\bbP^1\times \bbP^1$ serve as the  indeterminacy set for $i_1$ and for $i_2$. The singularity confinement structure is as in \eqref{sing1}.

{\bf Remark.}
In what follows, we restrict ourselves to the case $\ka_1+\ka_2=0$. This restriction is not essential, but will allow us to shorten some of the formulas. Thus, from now on we set
\begin{equation}\label{dPE8 kappa restr}
\ka_1=-\ka, \quad \ka_2=\ka.
\end{equation} 
If, additionally, the points $z_i$ satisfy the condition
\begin{equation}\label{dPE8 zi restr}
z_{i+4}=-z_i,\quad i=1,\ldots,4,
\end{equation}
then the QRT involutions admits a symmetry $i_1=\sigma\circ i_2\circ \sigma$, where $\sigma(x,y)=(y,x)$, so that one can introduce the QRT root $f=i_1\circ \sigma=\sigma\circ i_2$, such that $F=f\circ f$.

\paragraph{3D Painlev\'e map.} As usual, we identify $\bbP^1 \times \bbP^1$ with the quadric $Q_0 = \{ X_1 X_2 - X_3 X_4 =0 \} \subset \bbP^3$ via $[X_1 : X_2 : X_3 : X_4] = [x: y: xy: 1]$. The points $s_i$ are lifted to 
$$
S_i=[a_i:b_i:a_ib_i:1].
$$
We declare $Q_\lambda$ to be spanned by $Q_0$ and $Q_\infty=P_\infty = (X_1-X_2)^2-2\ka^2(X_1+X_2)X_4$:
\begin{equation}\label{dPE8 Q lambda}
	Q_\lambda = \Big\{ X_1 X_2 - X_3 X_4 - \lambda \big((X_1-X_2)^2-2\ka^2(X_1+X_2)X_4\big)=0\Big\}.
\end{equation}
The base set of the pencil $Q_\lambda$ is a cuspidal space curve of degree 4, $\{X_1 X_2 - X_3 X_4 =0,\, (X_1-X_2)^2-2\ka^2(X_1+X_2)X_4 =0 \}$, with the cusp at $[0:0:1:0]$. This is a pencil of type (iii).

The matrix $M_\lambda$ of the quadratic form $Q_\lambda$:
\begin{equation}
M_\lambda=\begin{pmatrix}
-2\lambda & 1+2\lambda & 0  & 2\lambda\ka^2\\
1+2\lambda & -2\lambda & 0 & 2\lambda\ka^2 \\
0 & 0 & 0 & -1 \\
2 \lambda\ka^2 & 2\lambda\ka^2 & -1 & 0 
\end{pmatrix}
\end{equation}
The characteristic polynomial of the pencil $\{Q_\lambda\}$ is: $\Delta(\lambda)=\det(M_\lambda)=1+4\lambda$, so that ${\rm Sing}(Q_\lambda)=\{-\frac{1}{4},\infty\}$. We set
\begin{equation}\label{dPE8 unif}
\lambda=\frac{\nu^2-1}{4}, \quad \sqrt{\Delta(\lambda)}=\nu.
\end{equation}
Thus, $\lambda(\nu)=\lambda(-\nu)$, while $ \sqrt{\Delta(\lambda)}$ changes its sign as $\nu\mapsto-\nu$. This gives us a double cover of the original pencil branched at $\nu=0$, corresponding to $\lambda=-1/4$, and at $\nu=\infty$, corresponding to $\lambda=\infty$. The normalizing transformation of $Q_\lambda$ to the canonical form $Q_0$ can be found as follows: 
\begin{equation}\label{dPE8 pencil norm}
\begin{bmatrix} X_1\\X_2\\X_3\\X_4\end{bmatrix} = A_\nu \begin{bmatrix} Y_1 \\ Y_2 \\ Y_3 \\ Y_4\end{bmatrix}, 
\end{equation}
where
\begin{equation}
A_\nu=\begin{pmatrix} \frac{1}{2\nu}(\nu+1) & \frac{1}{2\nu}(\nu-1) & 0 &  0 \\[0.15cm]
\frac{1}{2\nu}(\nu-1) & \frac{1}{2\nu}(\nu+1) & 0 & 0 \\[0.15cm] 
\frac{\ka^2}{2}(\nu^2-1) & \frac{\ka^2}{2}(\nu^2-1) & 1 & 0 \\[0.15cm]
0 & 0 & 0 & 1 \end{pmatrix}.                               
 \end{equation}
 Indeed, one immediately verifies that 
 $$
 A_\nu^{\rm T} M_{\lambda(\nu)} A_\nu=M_0.
 $$
Now, we are in the position to derive a parametrization of the quadric $Q_{\lambda(\nu)}$:
\begin{equation}\label{dPE8 x to X}
\begin{bmatrix} X_1 \\ X_2 \\ X_3 \\ X_4 \end{bmatrix}=A_\nu \begin{bmatrix} x \\ y \\ xy \\ 1 \end{bmatrix}
=\begin{bmatrix} \frac{1}{2\nu}((\nu+1)x+(\nu-1)y) \\[0.15cm]
 \frac{1}{2\nu}((\nu-1)x+(\nu+1)y) \\[0.15cm]
  xy+\frac{\ka^2}{2}(\nu^2-1)(x+y) \\[0.15cm] 1  \end{bmatrix}=:\phi_\nu(x,y).
\end{equation}
Observe that this parametrization is neither valid for $\nu=0$ nor for $\nu=\infty$.
The pencil-adapted coordinates $(x,y,\nu)$ on (the double cover of) $\bbP^3$ are:
\begin{equation}\label{dPE8 X to x}
x = \dfrac{(\nu+1)X_1-(\nu-1)X_2}{2X_4},\quad y = \dfrac{(\nu+1)X_2-(\nu-1)X_1}{2X_4},
\end{equation}
which have to be supplemented with
\begin{equation}\label{dPE8 X to lambda}
\lambda = \dfrac{\nu^2-1}{4} = \dfrac{X_1 X_2-X_3X_4}{(X_1-X_2)^2-2\ka^2(X_1+X_2)X_4}.
\end{equation}
The degenerate quadrics for $\nu = \infty$ and for $\nu=0$ are cones.

\begin{theorem}
For any $\delta\in\mathbb  C\setminus\{0\}$, define the Painlev\'e deformation map corresponding to the translation $\nu\mapsto\widehat\nu=\nu+2\delta$ by 
$$
L:\quad \left\{\begin{array}{ccl} \widehat X_1 & = & X_1X_4, \\ 
\widehat X_2 & = & X_2X_4, \\
\widehat X_3 & = & X_3X_4-\big(\lambda(\widehat\nu)-\lambda(\nu)\big)Q_\infty(X)\\
                      & = & X_3X_4- \beta(\nu+\beta)(X_1-X_2)^2+2\ka^2\beta (\nu+\beta)(X_1+X_2)X_4,\\
\widehat X_4 & = & X_4^2.
\end{array}\right.
$$
Then, in pencil-adapted coordinates, the map $L$ acts as follows:
\begin{equation}\label{dPE8 map L}
L: (x,y,\nu)\mapsto (\widehat x,\widehat y,\widehat \nu), \quad \widehat x =x+\frac{\delta(x-y)}{\nu},\quad \widehat y = y+\frac{\delta(y-x)}{\nu},\quad \widehat\nu=\nu+2\delta.
\end{equation}
For the latter map,  the factorizations \eqref{L fact 1}, \eqref{L fact 2} are given by
\begin{equation}\label{dPE8 L1}
L_1=R_1: (x,y,\nu)\mapsto (x,\widetilde y,\nu+\delta), \quad \t y= y+\dfrac{\delta}{\nu}(y-x),
\end{equation} 
\begin{equation}\label{dPE8 L2}
L_2=R_2:(x,y,\nu)\mapsto (\widetilde x,y, \nu+\delta),\quad \t x= x+\frac{\delta}{\nu}(x-y).
\end{equation}
\end{theorem}

\paragraph{Relation to the $d$-Painlev\'e equation of the surface type $A_0^{(1)}$.} 

In the pencil-adapted coordinates $(x,y,\nu)$, for each fixed $\nu$, the intersection curves $Q_{\lambda(\nu)} \cup P_\mu$ form the pencil through the points
\begin{equation}
s_i(\nu)=(a_i(\nu),b_i(\nu))=\big( z_i(z_i+\ka\nu), z_i(z_i-\ka\nu)\big), \quad i=1,\ldots,8,
\end{equation}
which are just the points $S_i$ expressed in the pencil-adapted coordinates on $Q_{\lambda(\nu)}$. Thus, the 3D QRT involutions $i_1$, $i_2$ act on each quadric $Q_{\lambda(\nu)}$ in the pencil-adapted coordinates via formulas which are obtained from the corresponding 2D formulas by replacing $\ka$ by $\ka \nu$:
\begin{align} \label{dPE8 3D QRT i1 on fiber}
& i_1(x,y)=(x,\t y), \nonumber\\  
& \frac{\big(\t y-\xi(\xi-\ka\nu)\big)\big(y-\xi(\xi-\ka\nu)\big)}{\big(\t y-(\xi+\ka\nu)(\xi+2\ka\nu)\big)\big(y-(\xi+\ka\nu)(\xi+2\ka\nu)\big)}=\frac{U(\xi)}{U(-\ka\nu-\xi)}, 
 \quad x=\xi(\xi+\ka\nu),
 \end{align}
 \begin{align} \label{dPE8 3D QRT i2 on fiber}
& i_2(x,y)=(\t x,y),\nonumber\\
&  \frac{\big(\t x-\eta(\eta+\ka\nu)\big)\big(x-\eta(\eta+\ka\nu)\big)}{\big(\t x-(\eta-\ka\nu)(\eta-2\ka\nu)\big)\big(x-(\eta-\ka\nu)(\eta-2\ka\nu)\big)}=\frac{U(\eta)}{U(\ka\nu-\eta)}, \quad y=\eta(\eta-\ka\nu).
 \end{align}
In notations of  \eqref{dP def eq 1}, \eqref{dP def eq 2}, the latter two equations take the following form:
 \begin{eqnarray} \label{dPE8 aux1}
\frac{\big(\t x-\eta(\eta+\ka\nu_{2n})\big)\big(x-\eta(\eta+\ka\nu_{2n})\big)}{\big(\t x-(\eta-\ka\nu_{2n})(\eta-2\ka\nu_{2n})\big)\big(x-(\eta-\ka\nu_{2n})(\eta-2\ka\nu_{2n})\big)}=\frac{U(\eta)}{U(\ka\nu_{2n}-\eta)},\nonumber\\ 
y_n=\eta(\eta-\ka\nu_{2n}), 
 \end{eqnarray}
\begin{eqnarray} \label{dPE8 aux2}
\frac{\big(\t y-\xi(\xi-\ka\nu_{2n+1})\big)\big(y-\xi(\xi-\ka\nu_{2n+1})\big)}{\big(\t y-(\xi+\ka\nu_{2n+1})(\xi+2\ka\nu_{2n+1})\big)\big(y-(\xi+\ka\nu_{2n+1})(\xi+2\ka\nu_{2n+1})\big)}=\frac{U(\xi)}{U(-\ka\nu_{2n+1}-\xi)},\nonumber\\
 x_{n+1}=\xi(\xi+\ka\nu_{2n+1}).\quad
 \end{eqnarray}
 Recall that here 
$$
\nu_{2n+1}=\nu_{2n+1/2}+\delta=\nu_{2n}+2\delta.
$$
To express in \eqref{dPE8 aux1} the variables $x$, $\t x$ through $x_n$, $y_n$, we observe that 
$$
L_2: (x_n,y_{n},\nu_{2n-1/2})\mapsto (x,y_n,\nu_{2n}), \quad R_2: (\t x,y_n,\nu_{2n})\mapsto (x_{n+1},y_n,\nu_{2n+1/2})
$$ 
can be written, according to \eqref{dPE8 L2}, as follows:
$$
x= x_n+\frac{\delta}{\nu_{2n-1/2}}(x_n-y_{n}),\quad {\rm resp.}\quad  x_{n+1}=\t x+\frac{\delta}{\nu_{2n}}(\t x-y_{n}).
$$
 A simple computation confirms that these relations are equivalent to
\begin{equation}\label{dPE8 aux 1.1}
\frac{x-\eta(\eta+\ka\nu_{2n})}{x-(\eta-\ka\nu_{2n})(\eta-2\ka\nu_{2n})}=
\frac{x_n-\eta(\eta+\ka\nu_{2n-1})}{x_n-(\eta-\ka\nu_{2n})(\eta-\ka\nu_{2n}-\ka\nu_{2n-1})}, \quad 
y_n=\eta(\eta-\ka\nu_{2n}),
\end{equation}
\begin{equation}\label{dPE8 aux 1.2}
\frac{\t x-\eta(\eta+\ka\nu_{2n})}{\t x-(\eta-\ka\nu_{2n})(\eta-2\ka\nu_{2n})}=
\frac{x_{n+1}-\eta(\eta+\ka\nu_{2n+1})}{x_{n+1}-(\eta-\ka\nu_{2n})(\eta-\ka\nu_{2n+1}-\ka\nu_{2n})}, \quad 
y_n=\eta(\eta-\ka\nu_{2n}).
\end{equation}
Similarly, to express in \eqref{dPE8 aux2} the variables $y$, $\t y$ through $x_{n+1}$, $y_n$, we observe that 
$$
L_1:(x_{n+1},y_n,\nu_{2n+1/2})\mapsto (x_{n+1},y,\nu_{2n+1}), \quad R_1:(x_{n+1},\t y,\nu_{2n+1})\mapsto (x_{n+1},y_{n+1},\nu_{2n+3/2}),
$$
which, according to \eqref{dPE8 L1}, can be put as follows:
$$
y=y_{n}+\frac{\delta}{\nu_{2n+1/2}}(y_{n}-x_{n+1}), \quad  y_{n+1}=\t y+\frac{\delta}{\nu_{2n+1}}(\t y-x_{n+1}).
$$
Again, these relations are equivalent to
\begin{eqnarray}\label{dPE8 aux 2.1}
\frac{y-\xi(\xi-\ka\nu_{2n+1})}{y-(\xi+\ka\nu_{2n+1})(\xi+2\ka\nu_{2n+1})}=
\frac{y_n-\xi(\xi-\ka\nu_{2n})}{y_n-(\xi+\ka\nu_{2n+1})(\xi+\ka\nu_{2n+1}+\ka\nu_{2n})}, \nonumber\\
\qquad x_{n+1}=\xi(\xi+\ka\nu_{2n+1})
\end{eqnarray}
\begin{eqnarray}\label{dPE8 aux 2.2}
\frac{\t y-\xi(\xi-\ka\nu_{2n+1})}{\t y-(\xi+\ka\nu_{2n+1})(\xi+2\ka\nu_{2n+1})}=
\frac{y_{n+1}-\xi(\xi-\ka\nu_{2n+2})}{y_{n+1}-(\xi+\ka\nu_{2n+1})(\xi+\ka\nu_{2n+2}+\ka\nu_{2n+1})}, \nonumber\\ 
x_{n+1}=\xi(\xi+\ka\nu_{2n+1}).
\end{eqnarray}
Substituting \eqref{dPE8 aux 1.1}--\eqref{dPE8 aux 2.2} into \eqref{dPE8 aux1}, \eqref{dPE8 aux2}, we arrive at the following system of non-autonomous difference equations for the variables $x_n, y_n$ :
 \begin{eqnarray} \label{dPE8 1}  
 \frac{\big(x_{n+1}-\eta(\eta+\ka\nu_{2n+1})\big)\big(x_n-\eta(\eta+\ka\nu_{2n-1})\big)}
{\big(x_{n+1}-(\eta-\ka\nu_{2n})(\eta-\ka\nu_{2n+1}-\ka\nu_{2n})\big)\big(x_n-(\eta-\ka\nu_{2n})(\eta-\ka\nu_{2n}-\ka\nu_{2n-1})\big)} \nonumber\\
=\frac{U(\eta)}{U(\ka\nu_{2n}-\eta)}, \quad y_n=\eta(\eta-\ka\nu_{2n}),
 \end{eqnarray}
 \begin{eqnarray} \label{dPE8 2}
 \frac{\big(y_{n+1}-\xi(\xi-\ka\nu_{2n+2})\big)\big(y_n-\xi(\xi-\ka\nu_{2n})\big)}
{\big(y_{n+1}-(\xi+\ka\nu_{2n+1})(\xi+\ka\nu_{2n+2}+\ka\nu_{2n+1})\big)\big(y_n-(\xi+\ka\nu_{2n+1})(\xi+\ka\nu_{2n+1}+\ka\nu_{2n})\big)}\nonumber\\
 =\frac{U(\xi)}{U(-\ka\nu_{2n+1}-\xi)}, 
 \quad x_{n+1}=\xi(\xi+\ka\nu_{2n+1}).\quad
 \end{eqnarray}
This is the $d$-Painlev\'e equation of the surface type $A_0^{(1)}$, as given in \cite{Y1}, \cite{KNY}.
\smallskip

{\bf Remark.} In the symmetric situation, when $U(z)=U(-z)$, the system \eqref{dPE8 1}, \eqref{dPE8 2} can be interpreted as a one-field second order difference equation, with $x_n=u_{2n-1}$ and $y_n=u_{2n}$. To see this, one should make  the change $\xi\mapsto -\xi$ in equation \eqref{dPE8 2}, after which it matches \eqref{dPE8 1}.

%%%%%%%%%%%%%%%%%%%%%%%%%%%%%%%%%%%%%%%
\section{From a pencil of type (ii) to the \textit{q}-Painlev\'e equation of the surface type $A_0^{(1)}$}
\label{sect qPE8}
%%%%%%%%%%%%%%%%%%%%%%%%%%%%%%%%%%%%%%%%

\begin{figure}[!ht]
\begin{center}
\begin{subfigure}[b]{0.3\textwidth}
         \centering
         \includegraphics[width=0.8\textwidth]{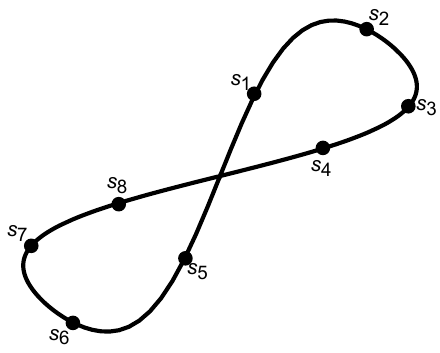}
         \caption{}
     \end{subfigure}
     \hspace{2cm}
     \begin{subfigure}[b]{0.3\textwidth}
         \centering
         \includegraphics[width=\textwidth]{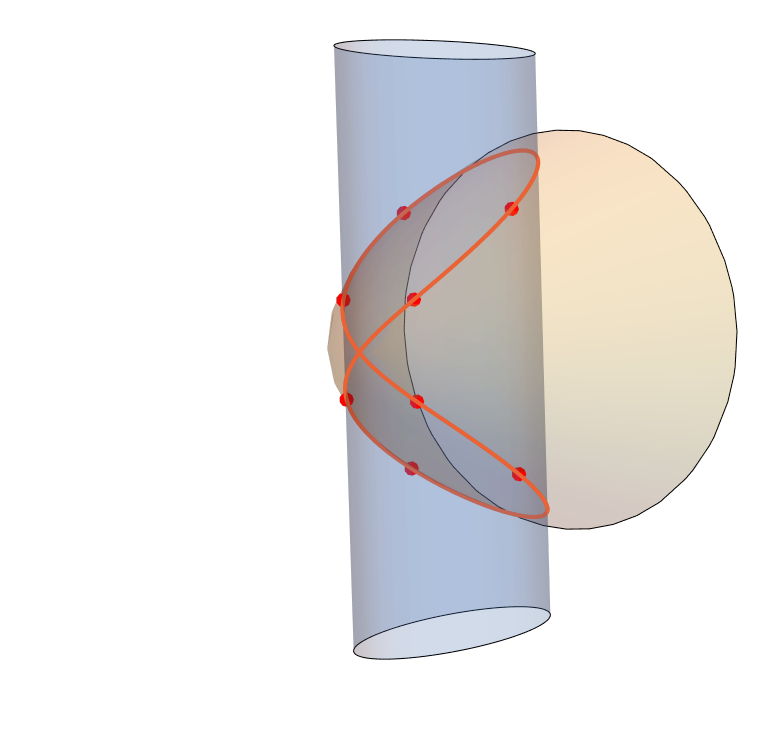}
         \caption{}
     \end{subfigure}
\end{center}
\caption{(a) Base set of the surface type $A_0^{(1)}$: eight points on a nodal (2,2)-curve in $\bbP^1\times\bbP^1$. (b) Pencil of quadrics through a nodal spatial quartic in $\bbP^3$}
\label{fig qPE8}
\end{figure}

\paragraph{2D QRT map.} We consider the QRT map corresponding to the pencil of biquadratic curves through eight points $s_i=(a_i,b_i)$, where
$$
a_i=z_i+\frac{\ka_1}{z_i}, \quad b_i=\frac{1}{z_i}+\frac{z_i}{\ka_2}, \quad i=1,\ldots,8.
$$
These eight points support a pencil of biquadratic curves if they satisfy the condition
$$
\prod_{i=1}^8 z_i=\ka_1^2\ka_2^2.
$$
They belong to the curve with the equation
$$
(x-\ka_2 y)(y-\ka_1^{-1} x)=(\ka_1\ka_2)^{-1}(\ka_1-\ka_2)^2.
$$
This is a biquadratic curve in $\bbP^1\times\bbP^1$ with a simple node at $(\infty,\infty)$, see Fig. \ref{fig qPE8} (a).

%%%%%%%%%%%%%%%%%%%%%%%%%%%%%%%%%
%%%%%%%%%%%%%%%%%%%%%%%%%%%%%%%%%

The vertical involution $i_1$ can be described by the following equation:
 \begin{equation} \label{qPE8 QRT i1}
 i_1(x,y)=(x,\t y),\quad
 \frac{\left(\t{y}-\dfrac{1}{\xi}-\dfrac{\xi}{\ka_2}\right)\left(y-\dfrac{1}{\xi}-\dfrac{\xi}{\ka_2}\right)}
{\left(\t{y}-\dfrac{\xi}{\ka_1}-\dfrac{\ka_1}{\ka_2\xi}\right)\left(y-\dfrac{\xi}{\ka_1}-\dfrac{\ka_1}{\ka_2\xi}\right)}
=\frac{U(\xi)}{U\Big(\dfrac{\ka_1}{\xi}\Big)}, \quad x=\xi+\frac{\ka_1}{\xi}.
 \end{equation}
 Here we use the abbreviation 
 \begin{equation}\label{qPE8 U}
 U(z)=z^{-4}\prod_{i=1}^8 (z-z_i).
 \end{equation}
Formula \eqref{qPE8 QRT i1} is understood as follows. Written as a Laurent polynomial in $\xi$, it is anti-symmetric with respect to $\xi\leftrightarrow \ka_1/\xi$. Upon division by $\xi-\ka_1/\xi$, the resulting Laurent polynomial is symmetric and therefore it can be actually expressed as a polynomial in $x=\xi+\ka_1/\xi$. This defines $i_1$ as a birational involution (its symmetry w.r.t. $y\leftrightarrow \t y$ is obvious).

Similarly, the horizontal involution $i_2$ can be described by the following equation:
 \begin{equation} \label{qPE8 QRT i2}
i_2(x,y)=(\t x,y),\quad
\frac{\left(\t{x}-\eta-\dfrac{\ka_1}{\eta}\right)\left(x-\eta-\dfrac{\ka_1}{\eta}\right)}
{\left(\t{x}-\dfrac{\ka_2}{\eta}-\dfrac{\ka_1\eta}{\ka_2}\right)\left(x-\dfrac{\ka_2}{\eta}-\dfrac{\ka_1\eta}{\ka_2}\right)}
=\frac{U(\eta)}{U\Big(\dfrac{\ka_2}{\eta}\Big)}, \quad y=\frac{1}{\eta}+\frac{\eta}{\ka_2}.
 \end{equation}

The eight points $s_1,\ldots,s_8$ in $\bbP^1\times \bbP^1$ serve as the  indeterminacy set for $i_1$ and for $i_2$. The singularity confinement structure is as in \eqref{sing1}. The QRT map $F$ is the composition of these two involutions, $F=i_1\circ i_2$. 
\smallskip

{\bf Remark.}
In what follows, we restrict ourselves to the case $\ka_1\ka_2=1$. This restriction is not essential, but will allow us to shorten some of the formulas. Thus, from now on we set in this section
\begin{equation}\label{qPE8 kappa restr}
\ka_1=\frac{1}{\ka}, \quad \ka_2=\ka.
\end{equation} 
If, additionally, the points $z_i$ satisfy the condition
\begin{equation}\label{qPE8 zi restr}
z_{i+4}=z_i^{-1},\quad i=1,\ldots,4,
\end{equation}
then the QRT involutions admits a symmetry $i_1=\sigma\circ i_2\circ \sigma$, where $\sigma(x,y)=(y,x)$, so that one can introduce the QRT root $f=i_1\circ \sigma=\sigma\circ i_2$, such that $F=f\circ f$. 

\paragraph{3D Painlev\'e map.} As usual, we identify $\bbP^1 \times \bbP^1$ with the quadric $Q_0 = \{ X_1 X_2 - X_3 X_4 =0 \} \subset \bbP^3$ via $[X_1 : X_2 : X_3 : X_4] = [x: y: xy: 1]$. The points $s_i$ are lifted to 
$$
S_i=[a_i:b_i:a_ib_i:1].
$$
We declare $Q_\lambda$ to be spanned by $Q_0$ and
\begin{equation}\label{qPE8 Q lambda}
Q_\infty=P_\infty = \ka (X_1^2 + X_2^2) -(1+\ka^2)X_1 X_2 +(\ka-\ka^{-1})^2X_4^2.
\end{equation}
The base set of the pencil $Q_\lambda$ is a nodal space curve $\{Q_0 =0,\, P_\infty=0 \}$ of degree 4,  with the node at $[0:0:1:0]$. This is a pencil of type (ii).

The matrix $M_\lambda$ of the quadratic form $Q_\lambda$:
\begin{equation}
M_\lambda=\begin{pmatrix}
-2\ka\lambda & 1+(1+\ka^2)\lambda & 0  & 0\\
1+(1+\ka^2)\lambda & -2\ka\lambda & 0 & 0 \\
0 & 0 & 0 & -1 \\
0 & 0  & -1 & -2(\ka-\ka^{-1})^2\lambda 
\end{pmatrix}.
\end{equation}
The characteristic polynomial of the pencil $\{Q_\lambda\}$ is: 
$$
\Delta(\lambda)=\det(M_\lambda)=\big(1+(1+\ka^2)\lambda\big)^2-4\ka^2\lambda^2=\big(1+(1+\ka)^2\lambda\big)\big(1+(1-\ka)^2\lambda\big),
$$ 
so that ${\rm Sing}(Q_\lambda)=\{-(1+\ka)^{-2},-(1-\ka)^{-2},\infty\}$. This polynomial is not a complete square, and we have to uniformize $\sqrt{\Delta(\lambda)}$. The uniformizing variable is $\nu\in\bbC$. As in Sect. \ref{sect qPE7}, it will be convenient to use $w=e^\nu$ instead, with $w\in \bbC\setminus \{0\}$.  We set
\begin{equation}\label{qPE8 lambda unif}
	\lambda =\lambda(w)= \dfrac{(\ka-w)(1-\ka w)}{(1-\ka^2)^2 w}.
\end{equation} 
Then $\Delta(\lambda)$ becomes a square:
$$
\Delta(\lambda)=\dfrac{\ka^2(1-w^2)^2}{w^2(1-\ka^2)^2}\quad \Rightarrow\quad \sqrt{\Delta(\lambda)}=\dfrac{\ka(1-w^2)}{w(1-\ka^2)}.
$$
Observe that $\lambda(w)=\lambda(w^{-1})$, while $\sqrt{\Delta(\lambda)}$ changes its sign under $w\mapsto w^{-1}$. This gives us a double cover of the original pencil branched at $\lambda=-(1+\ka)^{-2}$ (corresponding to $w=1$), and at $\lambda=-(1-\ka)^{-2}$ (corresponding to $w=-1$). The point $\lambda=\infty$ is not a branch point (it corresponds to $w=0,\infty$). The normalizing transformation of $Q_\lambda(X)$ to the canonical form $Q_0(Y)=Y_1Y_2-Y_3Y_4$ is achieved by the transformation
\begin{equation}\label{qPE8 pencil norm}
\begin{bmatrix} X_1\\X_2\\X_3\\X_4\end{bmatrix} = A_w \!\begin{bmatrix} Y_1 \\ Y_2 \\ Y_3 \\ Y_4\end{bmatrix}, 
\end{equation}
where one can take
\begin{equation}
A_w=\begin{pmatrix} 
 	\dfrac{w(1-\ka w)}{\ka(1-w^2)} & \dfrac{w(\ka-w)}{\ka(1-w^2)} & 0 & 0 \\[0.25cm] 
 	\dfrac{w(\ka-w)}{\ka(1-w^2)} & \dfrac{w(1-\ka w)}{\ka(1-w^2)} & 0 & 0 \\
 	0&0 & \dfrac{w}{\ka} & -\dfrac{(1-\ka w)(\ka-w)}{\ka^2 w} \\[0.25cm]
 	0 & 0 & 0 & 1
 	 \end{pmatrix}.                          
 \end{equation}
 Indeed, one immediately verifies that 
 $$
 A_w^{\rm T} M_{\lambda(w)} A_w=\frac{w}{\ka} M_0.
 $$
There follows a parametrization of the quadric $Q_{\lambda(w)}$:
\begin{equation}\label{qPE8 x to X}
\begin{bmatrix} X_1 \\ X_2 \\ X_3 \\ X_4 \end{bmatrix} =A_w
 	\begin{bmatrix} x \\ y \\ xy \\ 1 \end{bmatrix} =: \phi_w(x,y).
\end{equation}
This parametrization is neither valid for $w=0$ nor for $w=\infty$.
The pencil-adapted coordinates $(x,y,w)$ on (the double cover of) $\bbP^3$ are:
\begin{equation}\label{qPE8 X to x}
	x = \frac{\ka}{w}\cdot\dfrac{(1-\ka w)X_1-(\ka-w)X_2}{(1-\ka^2)X_4},\quad y = \frac{\ka}{w}\cdot\dfrac{(1-\ka w)X_2-(\ka-w)X_1}{(1-\ka^2)X_4},
\end{equation}
which have to be supplemented with
\begin{equation}\label{qPE8 X to lambda}
\lambda = \dfrac{(\ka-w)(1-\ka w)}{(\ka^2-1)^2 w} = \dfrac{X_1 X_2-X_3X_4}{\ka X_1^2 + \ka X_2^2 -(1+\ka^2)X_1 X_2 
+(\ka-\ka^{-1})^2X_4^2}.
\end{equation}

\begin{theorem}
For any $q\neq \pm 1$, define the Painlev\'e deformation map corresponding to the translation $w\mapsto \widehat w=q^2 w$ by 
\begin{equation}\label{qPE8 map L in X}
L: \quad\left\{ \begin{array}{l} \widehat X_1 = X_1X_4,\\
\widehat X_2=X_2X_4,\\
\widehat X_3=X_3X_4-\big(\lambda(\widehat w)-\lambda(w)\big)Q_\infty(X),\\
\widehat X_4=X_4^2,
\end{array}\right.
\end{equation}
where $\lambda=\lambda(w)$ is given by \eqref{qPE8 lambda unif}, and $Q_\infty(X)$ is given in \eqref{qPE8 Q lambda}. Then, in pencil-adapted coordinates, the map $L$ acts as follows:
\begin{equation}\label{qPE8 map L}
L: \quad \widehat x=x+\frac{1-q^{-2}}{w^2-1}(x-w y),\quad 
\widehat y=y+\frac{1-q^{-2}}{w^2-1}(y-w x), \quad \widehat w=q^2 w.
\end{equation}
For the latter map,  the factorizations \eqref{L fact 1}, \eqref{L fact 2} are given by
\begin{align}
& L_1: (x,y,w)\mapsto (x,\t  y, qw), \quad \t y= y+\frac{1-q^{-2}}{w^2-1}(y-qw x), 
\label{qPE8 L1}\\
& R_1: (x,y,w)\mapsto (x,\t y, qw), \quad \t y= y+\frac{1-q^{-2}}{w^2-1}(y-w x),
\label{qPE8 R1}\\
& L_2: (x,y,w)\mapsto (\t x, y, qw), \quad \t x= x+\frac{1-q^{-2}}{w^2-1}(x-qw y),
\label{qPE8 L2}\\
& R_2: (x,y,w)\mapsto (\t x, y, qw), \quad \t x= x+\frac{1-q^{-2}}{w^2-1}(x-w y).
\label{qPE8 R2}
\end{align}
\end{theorem}

\paragraph{Relation to the $q$-Painlev\'e equation of the surface type $A_0^{(1)}$.} 

In the pencil-adapted coordinates $(x,y,w)$, for each fixed $w$, the intersection curves $Q_{\lambda(w)} \cup P_\mu$ form the pencil through the points
\begin{equation}
s_i(w)=(a_i(w),b_i(w))=\Big( z_i+\frac{1}{w z_i}, \frac{1}{z_i}+\frac{z_i}{w}\Big), \quad i=1,\ldots,8,
\end{equation}
which are just the points $S_i$ expressed in the pencil-adapted coordinates on $Q_{\lambda(w)}$. Thus, the 3D QRT involutions $i_1$, $i_2$ act on each quadric $Q_{\lambda(w)}$ in the pencil-adapted coordinates via formulas which are obtained from the corresponding 2D formulas by replacing $\ka$ by $w$:
\begin{equation} \label{qPE8 3D QRT i1 on fiber}
 i_1(x,y)=(x,\t y),\quad
 \frac{\left(\t{y}-\dfrac{1}{\xi}-\dfrac{\xi}{w}\right)\left(y-\dfrac{1}{\xi}-\dfrac{\xi}{w}\right)}
{\left(\t{y}-w\xi-\dfrac{1}{w^2\xi}\right)\left(y-w\xi-\dfrac{1}{w^2\xi}\right)}
=\frac{U(\xi)}{U\Big(\dfrac{1}{w\xi}\Big)}, \quad x=\xi+\frac{1}{w\xi},
\end{equation}
 \begin{equation} \label{qPE8 3D QRT i2 on fiber}
i_2(x,y)=(\t x,y),\quad
\frac{\left(\t{x}-\eta-\dfrac{1}{w\eta}\right)\left(x-\eta-\dfrac{1}{w\eta}\right)}
{\left(\t{x}-\dfrac{w}{\eta}-\dfrac{\eta}{w^2}\right)\left(x-\dfrac{w}{\eta}-\dfrac{\eta}{w^2}\right)}
=\frac{U(\eta)}{U\Big(\dfrac{w}{\eta}\Big)}, \quad y=\frac{1}{\eta}+\frac{\eta}{w}.
 \end{equation}
In notations of \eqref{dP def eq 1}, \eqref{dP def eq 2}, this takes the form
 \begin{equation} \label{qPE8 aux1}
 \frac{\left(\t{x}-\eta-\dfrac{1}{w_{2n}\eta}\right)\left(x-\eta-\dfrac{1}{w_{2n}\eta}\right)}
{\left(\t{x}-\dfrac{w_{2n}}{\eta}-\dfrac{\eta}{w_{2n}^2}\right)\left(x-\dfrac{w_{2n}}{\eta}-\dfrac{\eta}{w_{2n}^2}\right)}
=\frac{U(\eta)}{U\Big(\dfrac{w_{2n}}{\eta}\Big)}, \quad
y_n=\frac{1}{\eta}+\frac{\eta}{w_{2n}}, 
 \end{equation}
\begin{equation} \label{qPE8 aux2}
 \frac{\left(\t{y}-\dfrac{1}{\xi}-\dfrac{\xi}{w_{2n+1}}\right)\left(y-\dfrac{1}{\xi}-\dfrac{\xi}{w_{2n+1}}\right)}
{\left(\t{y}-w_{2n+1}\xi-\dfrac{1}{w_{2n+1}^2\xi}\right)\left(y-w_{2n+1}\xi-\dfrac{1}{w_{2n+1}^2\xi}\right)}
=\frac{U(\xi)}{U\Big(\dfrac{1}{w_{2n+1}\xi}\Big)}, \quad x_{n+1}=\xi+\frac{1}{w_{2n+1}\xi}. 
\end{equation}
Here, recall,
\begin{equation}
w_{2n+1}=q w_{2n+1/2}=q^2 w_{2n}.
\end{equation}
To express in \eqref{qPE8 aux1} the variables $x$, $\t x$ through $x_n$, $y_n$, we observe that 
$$
L_2:(x_n,y_{n},w_{2n-1/2})\mapsto (x,y_n,w_{2n}), \quad
R_2:(\t x,y_n,w_{2n})\mapsto (x_{n+1},y_n,w_{2n+1/2}).
$$ 
According to \eqref{qPE8 L2}, \eqref{qPE8 R2}, we find:
$$
x=x_n+\frac{1-q^{-2}}{w_{2n-1/2}^2-1} (x_n-qw_{2n-1/2}y_n), \quad x_{n+1}=\t x+\frac{1-q^{-2}}{w_{2n}^2-1}(\t x -w_{2n} y_n).
$$
A straightforward computation confirms that these equations are equivalent to
\begin{equation}\label{qPE8 aux 1.1}
 \frac{x-\eta-\dfrac{1}{w_{2n}\eta}}
{x-\dfrac{w_{2n}}{\eta}-\dfrac{\eta}{w_{2n}^2}}
= \frac{x_n-\eta-\dfrac{1}{w_{2n-1}\eta}}
{x_n-\dfrac{w_{2n}}{\eta}-\dfrac{\eta}{w_{2n}w_{2n-1}}}, \quad y_n=\frac{1}{\eta}+\frac{\eta}{w_{2n}},
\end{equation}
\begin{equation}\label{qPE8 aux 1.2}
 \frac{\t{x}-\eta-\dfrac{1}{w_{2n}\eta}}
{\t{x}-\dfrac{w_{2n}}{\eta}-\dfrac{\eta}{w_{2n}^2}}
= \frac{x_{n+1}-\eta-\dfrac{1}{w_{2n+1}\eta}}
{x_{n+1}-\dfrac{w_{2n}}{\eta}-\dfrac{\eta}{w_{2n}w_{2n+1}}}, \quad y_n=\frac{1}{\eta}+\frac{\eta}{w_{2n}}.
\end{equation}
Similarly, to express in \eqref{qPE8 aux2} the variables $y$, $\t y$ through $x_{n+1}$, $y_n$, we observe that 
$$
L_1: (x_{n+1},y_n,w_{2n+1/2})\mapsto (x_{n+1},y,w_{2n+1}), \quad
R_1: (x_{n+1},\t y,w_{2n+1})\mapsto (x_{n+1},y_{n+1},w_{2n+3/2}).
$$
According to \eqref{qPE8 L1}, \eqref{qPE8 R1}, we find:
$$
y=y_n+\frac{1-q^{-2}}{w_{2n+1/2}^2-1} (y_n-qw_{2n+1/2}x_{n+1}), \quad 
y_{n+1}=\t y+\frac{1-q^{-2}}{w_{2n+1}^2-1}(\t y-w_{2n+1} x_{n+1}).
$$
These equations are equivalent to 
\begin{equation}\label{qPE8 aux 2.1}
 \frac{y-\dfrac{1}{\xi}-\dfrac{\xi}{w_{2n+1}}}
{y-w_{2n+1}\xi-\dfrac{1}{w_{2n+1}^2\xi}}=
 \frac{y_n-\dfrac{1}{\xi}-\dfrac{\xi}{w_{2n}}}
{y_n-w_{2n+1}\xi-\dfrac{1}{w_{2n+1}w_{2n}\xi}}, \quad x_{n+1}=\xi+\frac{1}{w_{2n+1}\xi},
\end{equation}
\begin{equation}\label{qPE8 aux 2.2}
 \frac{\t{y}-\dfrac{1}{\xi}-\dfrac{\xi}{w_{2n+1}}}
{\t{y}-w_{2n+1}\xi-\dfrac{1}{w_{2n+1}^2\xi}}=
 \frac{y_{n+1}-\dfrac{1}{\xi}-\dfrac{\xi}{w_{2n+2}}}
{y_{n+1}-w_{2n+1}\xi-\dfrac{1}{w_{2n+2}w_{2n+1}\xi}}, \quad x_{n+1}=\xi+\frac{1}{w_{2n+1}\xi}.
\end{equation}
Substitute \eqref{qPE8 aux 1.1}--\eqref{qPE8 aux 2.2} into \eqref{qPE8 aux1}, \eqref{qPE8 aux2}. This results in the following system of non-autonomous difference equations for the variables $x_n, y_n$:
 \begin{equation} \label{qPE8 1}
 \frac{\left(x_{n+1}-\eta-\dfrac{1}{w_{2n+1}\eta}\right)\left(x_n-\eta-\dfrac{1}{w_{2n-1}\eta}\right)}
{\left(x_{n+1}-\dfrac{w_{2n}}{\eta}-\dfrac{\eta}{w_{2n}w_{2n+1}}\right)\left(x_n-\dfrac{w_{2n}}{\eta}-\dfrac{\eta}{w_{2n}w_{2n-1}}\right)}
=\frac{U(\eta)}{U\Big(\dfrac{w_{2n}}{\eta}\Big)}, \quad
y_n=\frac{1}{\eta}+\frac{\eta}{w_{2n}}, 
 \end{equation}
\begin{eqnarray} \label{qPE8 2}
 \frac{\left(y_{n+1}-\dfrac{1}{\xi}-\dfrac{\xi}{w_{2n+2}}\right)\left(y_n-\dfrac{1}{\xi}-\dfrac{\xi}{w_{2n}}\right)}
{\left(y_{n+1}-w_{2n+1}\xi-\dfrac{1}{w_{2n+2}w_{2n+1}\xi}\right)\left(y_n-w_{2n+1}\xi-\dfrac{1}{w_{2n+1}w_{2n}\xi}\right)}
=\frac{U(\xi)}{U\Big(\dfrac{1}{w_{2n+1}\xi}\Big)}, \nonumber\\
x_{n+1}=\xi+\frac{1}{w_{2n+1}\xi}. 
\end{eqnarray}
This is the $q$-Painlev\'e equation of the surface type $A_0^{(1)}$, as given in \cite{Y1}, \cite{KNY}.
\smallskip

{\bf Remark.} In the symmetric situation, when $U(z)=U(z^{-1})$, the system \eqref{qPE8 1}, \eqref{qPE8 2} can be interpreted as a one-field second order difference equation, with $x_n=u_{2n-1}$ and $y_n=u_{2n}$. To see this, one should make in equation \eqref{qPE8 2} the change $\xi\mapsto \xi^{-1}$, after which it matches \eqref{qPE8 1}.

%%%%%%%%%%%%%%%%%%%%%%%%%%%%%%%%%%%%%%%%%%%%%%

\section{Conclusions}

In this paper, we carried out the largest part of the task left open in \cite{ASWpart1}, namely extended our novel approach to the pencils for which the generators through a point $X\in\mathbb P^3$ depend on $X$ in a non-rational (branching) way. The only case left open for a further investigation is the pencil of the generic type (i), associated (in our scheme) with the elliptic Painlev\'e equation.  Also the problem of an interpretation of the isomonodromic property of discrete Painlev\'e equations within our scheme remains open and is left for the future research.

\end{document}